\documentclass{article}

\usepackage{PRIMEarxiv}

\usepackage[utf8]{inputenc} 
\usepackage[T1]{fontenc}    
\usepackage{hyperref}       
\usepackage{url}            
\usepackage{booktabs}       
\usepackage{amsfonts}       
\usepackage{nicefrac}       
\usepackage{microtype}      
\usepackage{lipsum}
\usepackage{amsmath}
\usepackage{fancyhdr}       
\usepackage{graphicx}       
\graphicspath{{media/}}     

\usepackage{graphicx}

\usepackage{txfonts}

\usepackage[labelfont=bf,justification=justified]{caption}
\DeclareCaptionFormat{cont}{#1\textbf{.continued}#2#3\par}
\usepackage{indentfirst}
\usepackage[title]{appendix}
\usepackage[export]{adjustbox}
\usepackage{float}
\usepackage{hyperref}
\usepackage{natbib}
\newcommand{\gdor}{$\gamma$~Dor}
\newcommand{\ds}{$\delta$~Sct}
\newcommand{\dss}{$\delta$~Sct stars}

\newcommand{\logg}{$\log g$}

\newcommand{\kms}{$\mathrm{km}~\mathrm{s}^{-1}$}
\newcommand{\teff}{T$_{\mathrm{eff}}$}

\newcommand{\cd}{$\mbox{d}^{-1}$}
\newcommand{\ele}{$\ell$}

\newcommand{\multim}{{\sc MultiModes}}
\newcommand{\mesa}{{\sc mesa}}
\newcommand{\gyre}{{\sc gyre}}
\newcommand{\myr}{$\mathrm{Myr}$}

\title{Comparative analysis of two $\delta$ Scuti stars approaching the TAMS, using MESA-GYRE models 
}

\author{
  David Pamos Ortega \\
  IES Levante \\
  Departamento de Física y Química \\
  Algeciras (Cádiz)\\
  \texttt{david.pamos.or@gmail.com} 
}

\begin{document}
\maketitle

\begin{abstract}
This study presents a comparative asteroseismic analysis of two late main-sequence \dss\ with slow to moderate rotation, TIC~274211446 and TIC~16053262. Both are also characterised by showing an intense emission in the H$_{\alpha}$ line.\

Based on the first 35 extracted frequencies from short-cadence photometric data from the TESS mission, with no significant pulsations out of the [9, 40] \cd range, I estimated the large frequency separation using the Fourier transform, the autocorrelation and the histogram of frequency differencies. I calibrated evolutionary and adiabatic oscillation models using the \mesa\ and \gyre\ codes under the Traditional Approximation of Rotation (TAR). Assuming a lack of visible $g$-modes in both cases, I computed only $p$-modes with radial orders ranging from n = 1 to n = 8, spherical degrees from \ele = 0 to \ele = 2, and azimuthal orders from $m$ = -2 to $m$ = 2. I calculated the low order large separation of the models by averaging across the different degrees and azimuthal orders. I used the observed large separation and spectroscopic temperature to constrain the models. Finally, I have used kernel curve fitting (KDE) to find the best profiles for both stars.\

With a hydrogen fraction of 21$\%$ in the convective core, TIC~274211446 is more evolved than TIC~16053262, which still retains 30$\%$ of its core fuel. The former has a large separation $\Delta\nu$ = (2.7$\pm$0.1) \cd, and points to a model with M = (1.84$\pm$0.07) M$_{\odot}$, a radius of R = (2.67$\pm$0.07) R$_{\odot}$, and an effective temperature of \teff = (7308$\pm$71) K. The latter, with $\Delta\nu$ = (3.25$\pm$0.1) \cd, points to a more compact model with M = (1.72$\pm$0.07)M$_{\odot}$, R = (2.28$\pm$0.05)R$_{\odot}$, and \teff = (7298$\pm$59) K. The equatorial rotation velocities are (28$\pm$1) \kms\ and (47$\pm$13) \kms, respectively. TIC~274211446 has an age of (1070$\pm$127) \myr, and TIC~16053262 is (1145$\pm$165) \myr\ old.\

These models have limitations related to the physics of angular momentum transport. The core of both stars is inaccessible, given that no $g$-mode multiplets are observed. It would allow an estimation of the internal rotation gradient. In addition, determining the spectroscopic projected rotation speed is useful to obtain their angles of inclination, and explain the differences in mode visibility.
\end{abstract}


\section{Introduction}\label{sec:introduction}

Between the main sequence (MS) and the instability strip \ds\ are found. They are Population I stars with spectral types between A2 and F2 and masses ranging from 1.5 to 2.5 M$_{\odot}$ \citep{Breger2000,RodriguezBreger2001,Aerts2010}. They can show dozens of radial and non-radial pulsation modes, with periods spanning from 15 minutes to 8 hours \citep{Breger2000,Uytterhoeven2011,Holdsworth2014,Guzik2021}. Not all of them are acoustic ($p$-modes); buoyancy modes ($g$-modes) have also been detected in \ds\ \citep{Uytterhoeven2011, Balona2015}. With space-based data, primarily from Kepler \citep{Borucki2010} and TESS \citep{Ricker}, the number of known \ds\ and \gdor\ has increased from a few hundred to thousands. Since the initial data provided by the CoRoT mission \citep{Baglin2006}, periodicities have been observed in the pulsation spectra of \dss\ \citep{GH2013,GH2015,GH2017,Paparo2016, Michel2017}. The large frequency separation, common to the asymptotic regime in solar-like stars \citep{Tassoul1980}, has also become visible in \dss, particularly in younger individuals \citep{Bedding2020}. Its scaling relation with stellar density \citep{Suarez2014,GH2015,GH2017} makes it a valuable parameter for constraining models. These authors also demonstrated that this relationship is independent of stellar rotation. However, estimating this parameter in stars that have evolved off the main sequence is difficult, primarily for two reasons: first, the avoided-crossing between $p$-modes and $g$-modes; and second, the splitting of non-radial modes due to rotation, which is typically moderate to high in these stars. 

Regarding avoided-crossings, I am going to describe it based on \cite{Aerts2010}. The Brunt-Väisälä frequency, $N$, is a measure of the stability of the internal gas layers of a star against convective movements. It is given by the Eq.~\ref{eq:BruntV}:
\begin{equation}
    N^{2}=\frac{g}{H_{P}}(\nabla_{ad}\ -\ \nabla\ +\ \chi_{\mu}\nabla_{\mu})
    \label{eq:BruntV},
\end{equation}
where $g$ is the local gravity, H$_{P}$ = -P/(dP/dr) is the pressure-height scale, $\nabla$ = (dlnT/dlnP) is the real thermal gradient, $\nabla_{ad}$ is the adiabatic thermal gradient, $\nabla_{\mu}$=(dln$\mu$/dlnP) is the mean molecular weight, $\chi_{\mu}=(\partial\ln\rho/\partial\ln\mu)$ is the sensitivity of the equation of state to changes in chemical composition, and $\rho$, P are, respectively, the local density and pressure in the stellar interior.

Within the convective core, the plasma is homogeneously mixed ($\nabla_{\mu}$=0), and the thermal gradient is nearly adiabatic ($\nabla\approx\nabla_{ad}$). Then $N^{2}\leq0$ and the $g$-modes cannot propagate here. At the boundary, as one transitions from the convective zone to the radiative zone, a very steep gradient or a discontinuous jump occurs, depending on whether or not overshooting is active. This creates a high, narrow peak in $N$, a composition bump. Outside the core, in the radiative zone, the composition stabilizes ($\nabla_{\mu}$=0), but transport is radiative ($\nabla<\nabla_{ad}$), so $N^{2}$ becomes positive. This is where $g$-modes are born and propagate.

On the other hand, the parameter that best characterizes non-radial acoustic modes is the Lamb frequency $S_{\ell}$, the minimum frequency at which an acoustic wave can propagate horizontally at a given depth within the star. It is given by the Eq.~\ref{eq:Lamb}:
\begin{equation}
    S_{\ell}=\frac{\ell(\ell+1)c_{s}^{2}}{r^{2}}
    \label{eq:Lamb},
\end{equation}
where \ele\ is the spherical grade, $c_{s}$ is the local sound speed and $r$ is the local radii. It depends critically on $l$ and $r$. The higher the spherical degree \ele, the more difficult it is for the wave to penetrate deep into the star. As $r\rightarrow0$, $S_{\ell}\rightarrow\infty$. For this reason, acoustic waves never reach or cross the center of the star. The high Lamb frequency acts as a mirror, reflecting them back outward.

In order to separate $p$-modes from $g$-modes, one must compare the calculated frequencies, $\omega$, with these two limits, $N$ and $S_{\ell}$, throughout the radii of the star. When $\omega<N$ and $\omega<S_{\ell}$, the mode behaves like a buoyancy wave. It reaches its maximum amplitude near the convective core. When $\omega>N$ and $\omega>S_{\ell}$, the mode behaves like an acoustic wave, reaching its maximum amplitude in the outer layers. If $N<\omega<S_{\ell}$ or $S_{\ell}<\omega<N$, the wave cannot propagate, and its amplitude decreases exponentially.

As stars evolve along the MS, after having consummed a large fraction of the available hydrogen and approaching the TAMS, $N$ increases. In this phase of stellar evolution, both types couple to produce mixed modes pulsations. They behave as $g$-modes deep within the star and as $p$-modes when they reflect off the core and propagate back towards the surface \citep{Dziembowski1997,Nowakowski2003,Dupret2005,Aerts2010}.

Secondly, these objects are typically fast rotators \citep{Royer2007}. As their rotation velocity increases, non-radial modes split into the azimuthal order $m$ according to $\omega_{n,\ell,m}$ = $\omega_{n,\ell}$ + $m$(1 - C$_{n,\ell}$)$\Omega$), where $\Omega$ is the rotation frequency and $C_{n,\ell}$ is the Ledoux constant, accounting for the influence of the Coriolis force. For slow rotators, the Ledoux constant is approximately zero, and the multiplets are symmetric around the central $m=0$ frequency \citep{Aerts2010}. However, the scenario becomes more complex in moderately and rapidly rotating stars, where multiplet symmetry breaks down \citep{Mirouh2022,Zhao2024}.

In summary, the presence of mixed modes and densely interleaved multiplets within the acoustic regime constitute the two primary reasons why mode identification is challenging in late-MS \dss, difficulting a precise characterisation of them using 1D models \citep{PamosOrtega2025}.

This is a comparative study of two similar F0-type stars with effective temperatures around 7300 K and a surface gravity around 3.9, using 1D \mesa/\gyre\ models. This is possible because both of them are not rapid rotators, and their spectra appear to be free of $g$- and mixed modes. Both stars are also notable for being emitters in the H$_{\alpha}$ line.

The paper is structured as follows: Sect.~\ref{sec:targets} introduces the two analysed stars; Sect.~\ref{sec:seismic} explains the frequency analysis conducted for each target, which determines the large frequency separation and identifies several low-order radial modes; Sect.~\ref{sec:grid} details the properties of the computed \mesa/\gyre\ grid and the statistical method for constraining the models; Sect.~\ref{sec:results} presents the discussion of the results; and finally, Sect.~\ref{sec:conclusions} lists the main conclusions of this work. 
   

\section{The targets}\label{sec:targets}

The stars studied here are TYC 67-812-1 and BD+14 1714. Hereafter, they are referred to by their identifier in the TESS catalogue \citep{Stassun2019}, TIC~274211446 and TIC~16053262, respectively. The key parameters are shown in Table~\ref{tb:Parameters_stars}. Below, in this paragraph, I present the astrometric and kinematic data obtained from Gaia DR3 \citep{Gaia2023, Lindegren2021}. TIC~274211446 is located in the Taurus constellation, at $\simeq$454 pc ($\simeq$1,479 light-years). It resides within the Local Arm (Orion Arm), pointing towards the outer region of the Milky Way. Three-dimensionally, it is positioned roughly three times deeper than the main core of the Taurus Molecular Cloud Complex ($\simeq$140 pc). Its radial velocity (+9.72 \kms) indicates a net recession caused by galactic differential rotation. Its proper motion (pmra = -5.665 mas/yr, pmdec = -14.976 mas/yr) translates to a tangential velocity of 75.9 \kms. With a total 3D space velocity of 76.5 \kms, it is classified as a mature field star in a slightly eccentric orbit. TIC~16053262 is located in the Gemini constellation, at $\simeq$270 pc ($\simeq$880 light-years), oriented towards the leading edge of the local disc rotation. It is in a low-density interstellar dust zone, near the outer boundaries of the Local Bubble. Its radial velocity is minimal (+2.41 \kms), demonstrating that it shares almost precisely the same orbit and velocity as the Sun. Its proper motion (pmra = -11.360 mas/yr, pmdec = -0.494 mas/yr) yields a tangential velocity of 53.9 km/s. Its 3D velocity of 54.0 \kms\ is typical for the thin disc, moving orderly and parallel to the galactic plane. The Renormalised Unit Weight Error (RUWE) is less than 1.4 in both cases (1.14 for TIC~274211446 and 0.97 for TIC~16053262), with the data fitting well to the solution of a single, isolated star. 

There are hardly any references to these stars in the literature. In the most recent study \citep{Zhang2023}, they used spectroscopic data from LAMOST to include them as candidates for magnetically active stars, due to their intense emission in the H$_{\alpha}$ line. The study demonstrated a correlation between the frequency of flare events and magnetic spots and the emission line intensity measured by LAMOST. The H$_{\alpha}$ line width values are very similar in both cases: EW(H$_{\alpha}$) = (-0.305$\pm$0.017) nm for TIC~274211446, and EW(H$_{\alpha}$) = (-0.292$\pm$0.019) nm for TIC~16053262, which overlap with the -0.3 nm critical threshold for formal classification as an emission-line star. The ratio between the minimum width and the maximum width on that line is 0.992 for TIC~274211446; it is not available for TIC~16053262. Since the emission is practically constant, they were ruled out as magnetically active stars. Alternatively, the strong emission in the H$_{\alpha}$ line can only be explained by the presence of a remanent circumstellar disk around both stars. FGK stars surrounded by a debris disk with ages exceeding one billion years have been reported \citep{Trilling2008}.  

\begin{table*}
    \centering
	\caption{Parameters of the studied \dss.}
    \renewcommand{\arraystretch}{1.5}
	\addtolength{\tabcolsep}{2.5 pt}
	\resizebox{\textwidth}{!}{  
    \begin{tabular}{cccccccccccccc}
			\hline
			TIC ID & RA (deg) & DE (deg) & p (mas)& D (pc) & T$_{mag}$ & E(B-V) & [Fe/H] & $\mathrm{T_{eff}}$ (K) & $\mathrm{log\ (L/L_{\odot})}$ & $\mathrm{M\ (M_{\odot})}$ & $\mathrm{R\ (R_{\odot})}$ & $\bar \rho\ (\bar \rho_{\odot})$ & \logg   \\
			\hline
	       	274211446 & 54.29$\pm$0.04 & 4.85$\pm$0.03 & 2.18$\pm$0.04 & 454$\pm$9 & 10.147$\pm$0.006 & 0.174$\pm$0.027 & 0.080$\pm$0.008 & 7309$\pm$136 & 1.19$\pm$0.04 & 1.65$\pm$0.27 & 2.45$\pm$0.14 & 0.11$\pm$0.04 & 3.88$\pm$0.09 \\
			16053262 & 114.54$\pm$0.04 & 14.29$\pm$0.04 & 3.67$\pm$0.04 & 270$\pm$3 & 8.840$\pm$0.007 & 0.016$\pm$0.004 & -0.009$\pm$0.009 & 7297$\pm$109 & 1.13$\pm$0.02 & 1.65$\pm$0.28 & 2.29$\pm$0.08 & 0.14$\pm$0.04 & 3.93$\pm$0.08 \\      
    \end{tabular}
    }
    
	\label{tb:Parameters_stars}
\end{table*}

\section{The frequency analysis}\label{sec:seismic}

\begin{figure}   
    \centering
    \includegraphics[width=8cm,height=10cm,keepaspectratio]{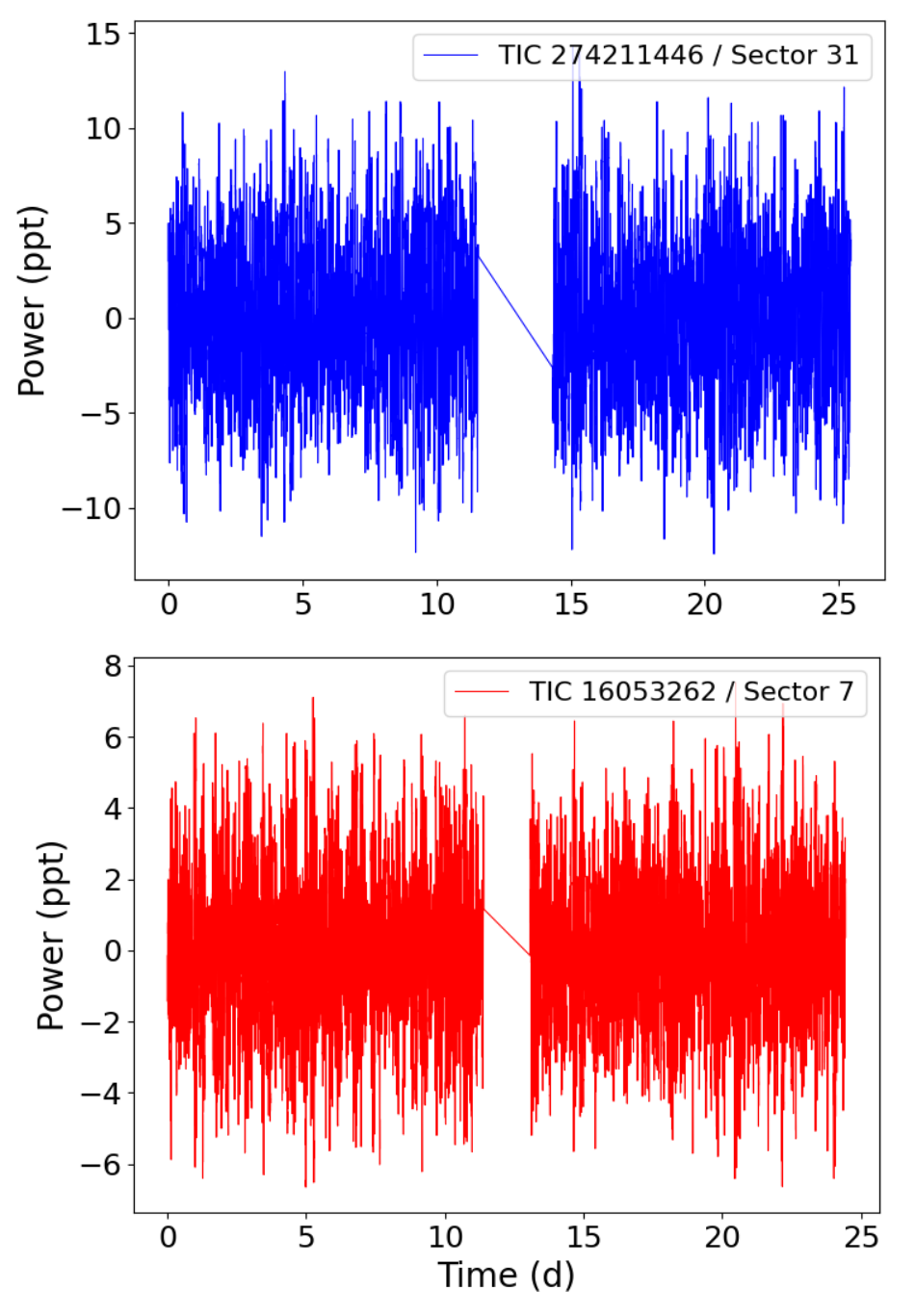}
    \caption{Light curves. Top: TIC~274211446. Bottom: TIC~16053262.}
    \label{fig:LC}
\end{figure}

Both stars were monitored by TESS at short cadence (120 s). TIC~274211446 was observed in sector 31 and TIC~16053262 in sector 7. I downloaded the PDCSAP$\_$FLUX light curves, which are corrected for instrumental effects, available in the Mikulski Archive for Space Telescopes,\footnote{\url{https://archive.stsci.edu}}. The Figure~\ref{fig:LC} show them. No signals can be seen that would allow us to infer any type of magnetic activity or atmospheric flares in both stars. 

\begin{figure}   
    \centering
    \includegraphics[width=11cm,height=13cm,keepaspectratio]{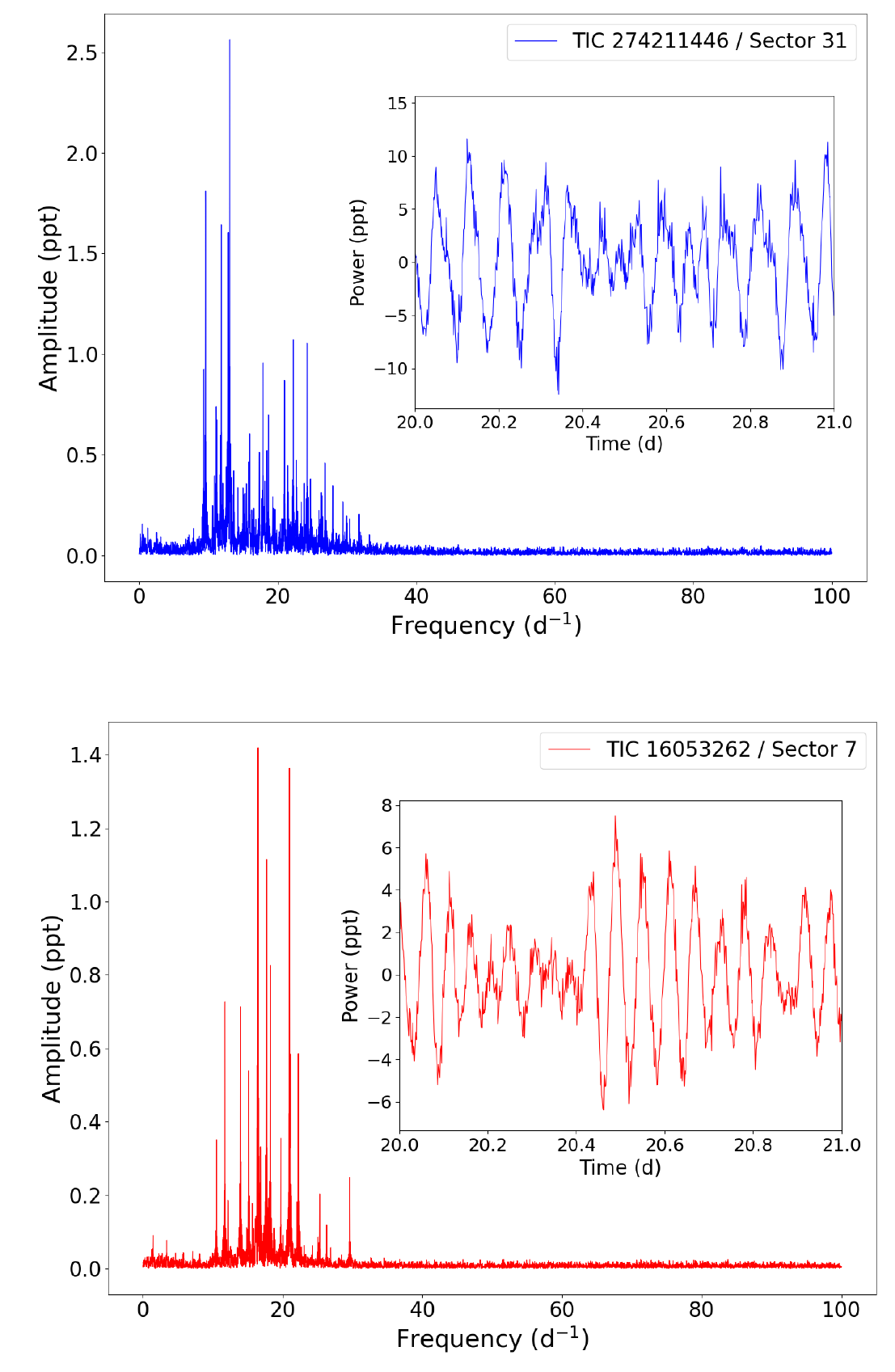}
    \caption{Light curves and periodograms. Left: TIC~274211446. Right: TIC~16053262.}
    \label{fig:LC_PG}
\end{figure}

I used \multim\footnote{\url{https://github.com/davidpamos/MultiModes}} for frequency extraction \citep{PamosOrtega2022}. The Figure~\ref{fig:LC_PG} shows the resulting periodograms. Zooming in on the light curves on day 21 of observation reveals a dominant pulsation with a frequency of approximately 13 \cd\ in TIC~274211446, whereas the dominant pulsation in TIC~16053262 has a frequency of around 16 \cd. Both periodograms are similar, with the significant frequency range spanning approximately between 9 and 30 \cd. This indicates a predominance of $p$-modes, albeit within a range where they may overlap with low order $g$- and mixed-modes. Despite being late-MS stars located near the red edge of the instability strip, I have analysed them as pure \dss, given that there is no presence of significant modes below 5 \cd, the boundary that separates them from \gdor\ stars \citep{Griga2010}. 

\begin{figure}   
    \centering
    \includegraphics[width=11cm,height=13cm,keepaspectratio]{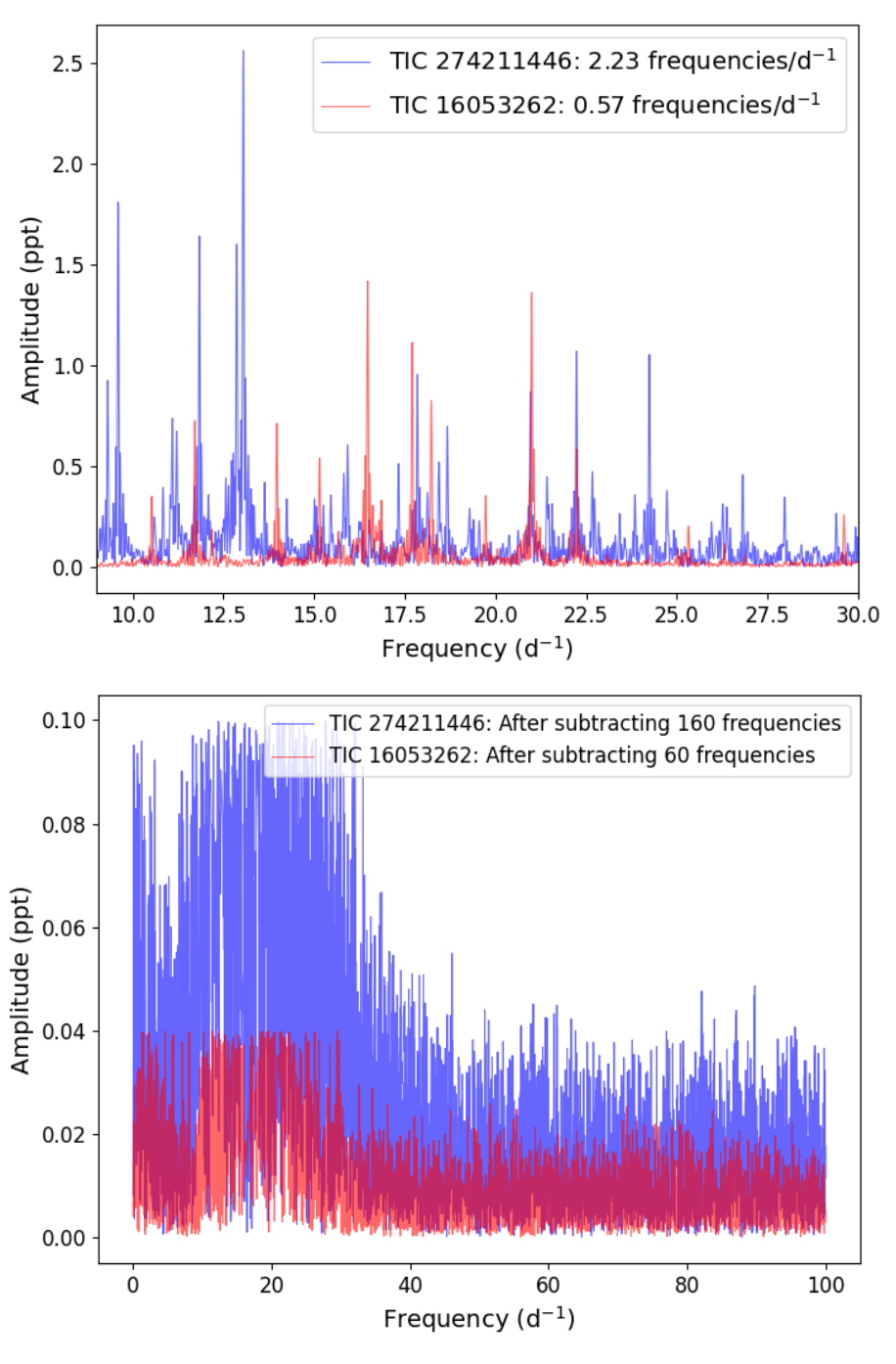}
    \caption{Top panel: Compared frequency density in both periodograms. Bottom panel: Comparison of the residuals following the extraction of significant frequencies (SNR > 4).}
    \label{fig:PG_compared}
\end{figure}

The upper panel of the Figure~\ref{fig:PG_compared} shows that TIC~274211446 has a periodogram nearly four times denser than that of TIC~16053262. Even although TIC~274211446 is more distant and is located in a more obscured environment than TIC~16053262, as confirmed by the higher reddening value, E(B-V), of the former. The lower panel of the Figure~\ref{fig:PG_compared} displays a plateau in the residual periodograms obtained upon completing the extraction of significant frequencies with signal to noise ratio of SNR > 4.0. This plateau is more pronounced in the star with the higher frequency density, as predicted by \cite{BF2024}. Furthermore, as explained by these authors, this feature serves as an indicator of the evolutionary stage of the star within the MS.

The Table~\ref{tb:Frequencies} shows the frequencies, amplitudes, phases, and SNRs of the first 35 extracted frequencies with \multim. Prior to the periodicity analysis, it was necessary to identify those modes that may be linear combinations of the highest-amplitude modes, adopting the Rayleigh resolution, approximately 0.041 \cd, as the tolerance threshold. Furthermore, for TIC~274211446, two peaks emerge that are suspected to be aliases of one of the highest-amplitude frequencies at 11.839 \cd. One of them, peak f$_{26}$ at 10.839 \cd, shows an exact shift of 1 \cd. The other, peak f$_{7}$ at 17.848 \cd, displays a shift of $\approx$6 c/d and shares the same phase value. Despite its considerable amplitude, ranking seventh in the list, I excluded it from the spectra periodicity analysis. Additionally, I discarded the f$_{27}$ peak that could represent a modulation of a neighbouring higher-amplitude peak, a common feature in these stars according to \cite{Bowman2016}. Regarding TIC~16053262, two modes of low amplitudes emerge with frequencies below 5 \cd. These two low amplitude modes, f$_{26}$ and f$_{29}$, were also excluded from the periodicity analysis. Finally, out of the first 35 extracted frequencies, 24 passed the selection criteria for TIC~274211446 and 29 for TIC~16053262.

\begin{figure*}   
    \centering
    \includegraphics[width=16cm,height=12cm,keepaspectratio]{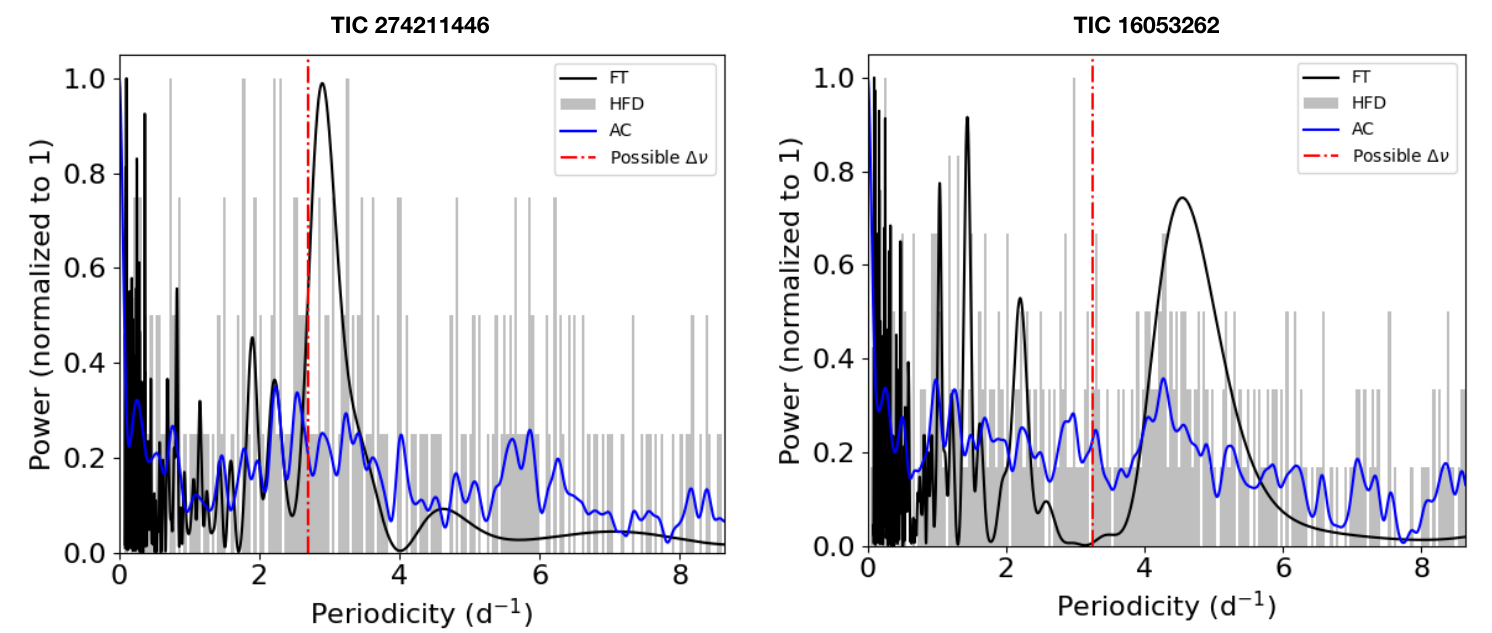}
    \caption{Large separation estimate through the Fourier transform(FT), the autocorrelation (AC) and the histogram of frequency differences (HFD). Left: TIC~274211446. Right: TIC~16053262.}
    \label{fig:dnus}
\end{figure*}

The Figure~\ref{fig:dnus} shows the analysis of the periodicities of the frequencies. I used the Fourier transform (FT), the autocorrelation (AC) and the histogram of frequency differences (HFD), following the techniques of \cite{GH2009, RBallesta2021}.  A prominent peak emerges around 2.7 \cd in the FT of TIC~274211446. It may be the large frequency separation, a value consistent with the star's evolutionary stage. It is not so obvious in the case of TIC~116053262. The FT shows two peaks, one around 2.2 \cd\ and the other around twice this value. None of these values are consistent with the star's evolutionary stage. Considering that it should be placed somewhere between the two peaks, I have estimated a value of 3.25 \cd\ for the second star. The Échelle diagrams (ED) do not help, at first glance, in determining the large frequency separation. Vertical alignments between modes of the same degree and consecutive orders are not readily apparent. However, as explained in Sect.~\ref{sec:results}, they become relevant when compared against models.

\section{The grid of models and the methodology}\label{sec:grid}

I used the MESA code, r-26.04.1 \citep{Paxton2011, Paxton2013, Paxton2019}, to compute evolutionary models up to the TAMS. The Table~\ref{tb:grid} shows the main grid parameters: masses ranging from 1.50 to 2.00 M$_{\odot}$, with a step size of 0.02 M$_{\odot}$. The metallicity, Z, ranges between 0.008 and 0.024, with a step size of 0.004. I have fixed the mixing-length parameter at $\alpha$ = 1.8 \citep{Dupret2005}. Overshooting was also included, as it is crucial at this evolutionary stage and key to determining the true age of the star, whose MS lifetime is extended by the injection of fresh hydrogen into the core. I have set values of $f$ = 0.015 and $f_{0}$ = 0.005 for the core, and $f$ = 0.002 and $f_{0}$ = 0.001 for the outer envelope \citep{Claret2019}. As in my previous work \citep{PamosOrtega2025}, where I computed models of \dss\ near the red edge of the instability strip, I have opted to retain the default boundary conditions for the Eddington atmosphere model. Rotation was activated at the ZAMS, with initial angular velocity values relative to the critical velocity ranging from 0.10 to 0.20, with a step size of 0.05. This allowed the stars to achieve equatorial rotation velocities between approximately 0 and 70 \kms. \mesa\ uses the shellular treatment of rotation \citep{Meynet2013}, which does not take into account the centrifugal deformation of the star. Although this 1D treatment breaks down at very high rotation rates, it is enough for the rotation rates discussed here. The transport of angular momentum is implemented via a diffusion approximation \citep{Paxton2013}. Regarding mesh resolution, I adopted the default spatial resolution value (\texttt{mesh\_delta\_coeff = 1.0}). However, I forced the code to implement finer time steps, particularly as it approaches the TAMS, by setting \texttt{time\_delta\_coeff} to 0.5. Consequently, changes in effective temperature and surface gravity are reduced, providing \gyre\ with a larger number of models at this stage for more precise frequency calculations.

\begin{table}
	\centering
	\caption{Parameters of the stellar models grid computed with the \mesa\ code.}
    \renewcommand{\arraystretch}{1.5}
	\addtolength{\tabcolsep}{8 pt}
	\resizebox{7cm}{!}{
    \begin{tabular}{ccc}
			\hline
			Parameter & Range & Step\\
			\hline
			M ({\rm M}$_{\odot}$) & [1.50; 2.00] & 0.02 ${\rm M}_\odot$\\
			Z & [0.008; 0.024] & 0.004\\
			$\Omega / \Omega_{c}$ & [0.10; 0.20] & 0.05\\
			$\alpha$ & 1.8 & Fixed\\
			
    \end{tabular}
    }
    
	\label{tb:grid}
\end{table}

Regarding the oscillation models, I used \gyre\ 9.1.1 \citep{Townsend2013}. Considering the frequency range in both stars, I computed only $p$-modes with radial orders ranging from n = 1 to n = 8, spherical degrees \ele = 0, 1, 2, and azimuthal orders between $m$ = -2 and $m$ = +2. Within the \texttt{\&OSC} namelist, I have adopted the standard Christensen-Dalsgaard outer boundary condition, \texttt{outer}$\_$\texttt{bound} = JCD and \texttt{variables}$\_$\texttt{bound} = JCD. I used the traditional approximation of rotation (TAR). For $p$-modes, the traditional perturbation methods have been shown to be reliable up to about $\Omega/\Omega_{c}$ = 0.4 \citep{Lignieres2006}. 

The numerical scheme for solving the equations was \texttt{MAGNUS}$\_$\texttt{GL2}, and a linear frequency scan (\texttt{grid}$\_$\texttt{type} = LINEAR) was employed. The frequency search window was restricted to values between 9 and 40 \cd, with a scan resolution of 200 points. Regarding the mesh resolution parameters in the \texttt{\&grid} namelist, I used the following values: \texttt{w}$_{osc}$ = 10, \texttt{w}$_{exp}$ = 2, \texttt{w}$_{ctr}$ = 10 and \texttt{n}$_{itermax}$= 5.

\begin{figure}   
    \centering
    \includegraphics[width=8cm,height=8cm,keepaspectratio]{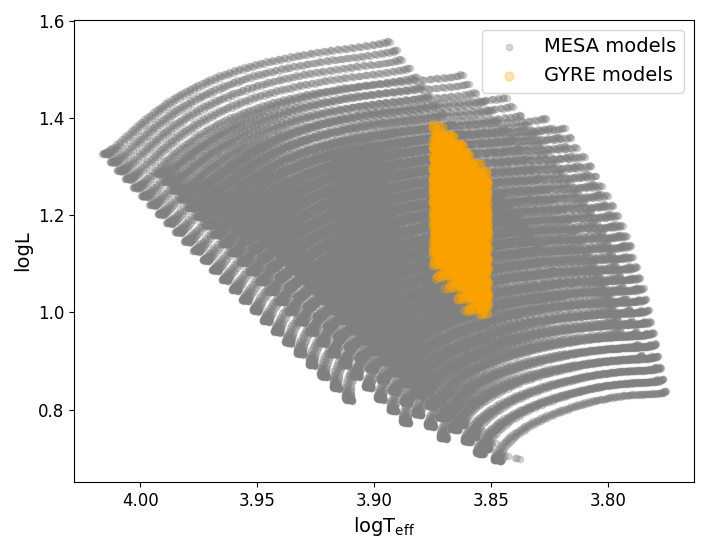}
    \caption{HR of the \mesa\ models. Highlighted in orange is the selection used to compute the oscillation models with \gyre.}
    \label{fig:HR_models}
\end{figure}

Once the model frequencies were computed, I calculated the large frequency separation of low orders (Eq.~\ref{eq:large_sep}) by averaging the differences between modes of consecutive radial orders across the different degrees and azimuthal orders, $\nu_{n\ell m}$:
\begin{equation}
    \Delta\nu_{n\ell m}=\nu_{n\ell m}\ -\ \nu_{n-1\ell m}
    \label{eq:large_sep}
\end{equation}
The Figure~\ref{fig:HR_models} shows the HR displaying the 28,126 computed \mesa\ models older than 20 \myr\ and with X$_{c}$ > 0.045, avoiding models near the ZAMS and very close to the TAMS. Highlighted in orange are 1,602 models that I selected to calculate the oscillation frequencies with \gyre, with effective temperatures and surface gravities within the uncertainty boxes of both stars (Table~\ref{tb:Parameters_stars}).

After estimating the large frequency separation, with the subsequent support of the ED and the physical consistency of the models (Sect.~\ref{sec:results}), I used it, along with the spectroscopic temperature, to constrain them. With the selected models, I plotted the histograms and the Kernel Density Estimation curves (KDE) to identify the best possible profile through a chi-squared minimisation process. As previously implemented in \citep{PamosOrtega2022, PamosOrtega2023,PamosOrtega2025}, I used a weighting factor to give greater weight to models with larger time steps, thereby compensating for the lower number of models produced in each evolutionary track. 

\section{Results}\label{sec:results}
As discussed in Sect.~\ref{sec:seismic}, estimating the large frequency separation was challenging, given that the EDs did not initially clarify the results obtained from the other three techniques. It remains necessary to provide clearer evidences to robustly support the estimation of the large frequency separation, upon which this study is based. This is only possible once I have the candidate models against which the observations can be compared.

\begin{figure*}   
    \centering
    \includegraphics[width=16cm,height=12cm,keepaspectratio]{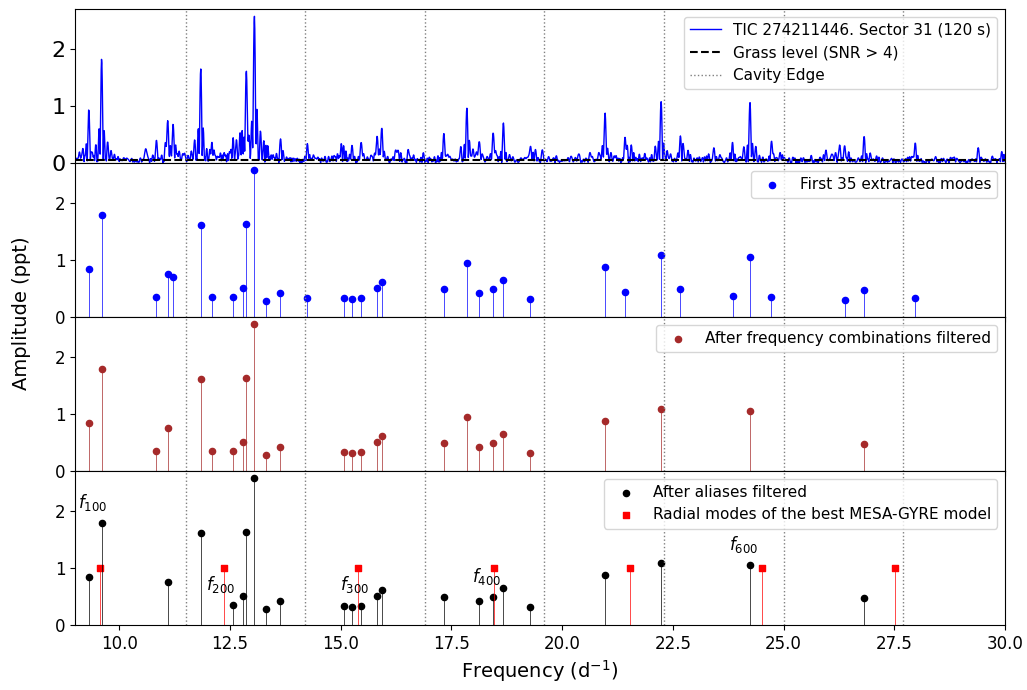}
    \caption{From top to bottom: Periodogram of TIC~274211446. The first 35 extracted frequencies with the highest amplitudes. Frequencies remaining after removing the linear combinations. Frequencies remaining after removing the detected aliases. The red squares represent the positions of the primary radial modes.}
    \label{fig:First35_TIC_1}
\end{figure*}

\begin{figure*}   
    \centering
    \includegraphics[width=16cm,height=12cm,keepaspectratio]{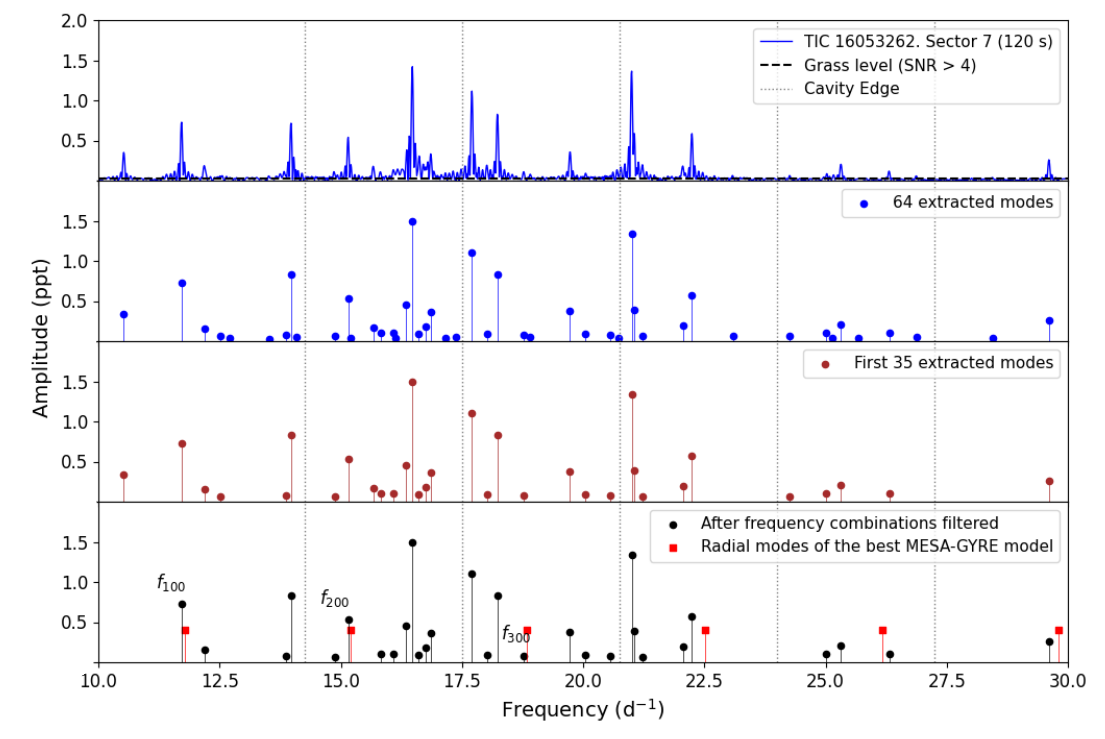}
    \caption{The same as Figure ~\ref{fig:First35_TIC_1} with TIC~16053262.}
    \label{fig:First35_TIC_2}
\end{figure*}

A visual inspection of the main extracted modes (Figure~\ref{fig:First35_TIC_1} and Figure~\ref{fig:First35_TIC_2}) has also contributed to estimating the large separation and identifying the primary radial modes in both stars. Starting with the 35 extracted highest-amplitude frequencies, I removed linear combinations as well as those suspected of being aliases or amplitude modulations of the primary mode (Sect.~\ref{sec:seismic}). The wider gaps between one group of frequencies and the next point to an estimate of the large separation, marked by a dotted grey line running vertically across all the panels in both figures. Furthermore, in order to compare the observations with the models, in the bottom panels I have indicated the low-order radial modes from the best-fitting models using red squares. 

\begin{figure*}   
    \centering
    \includegraphics[width=14cm,height=14cm,keepaspectratio]{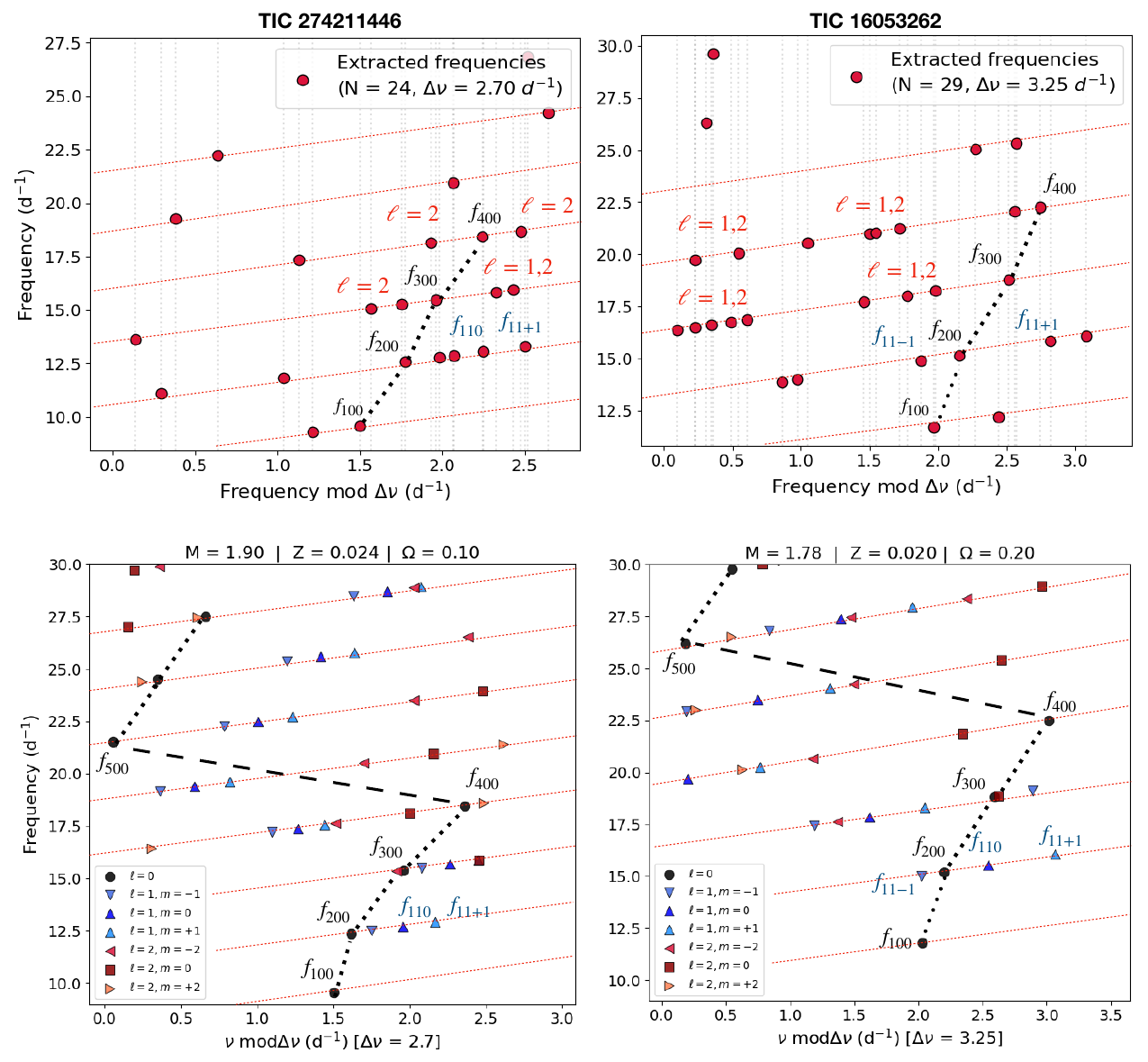}
    \caption{Top panel: Observed ED. Bottom panel: ED of the best model. Left: TIC~274211446. Right: TIC~16053262.}
    \label{fig:ED}
\end{figure*}

The Figure~\ref{fig:ED} illustrates the comparison between the observed ED (upper panels) and the model ED (lower panels), with TIC~274211446 on the left and TIC~16053262 on the right. I have highlighted some of the lowest-order radial modes, connected by a black dotted line. In order to assist in the visualisation and interpretation of the diagrams, for both the observed and model plots, the boundaries between consecutive orders have been represented by a red dotted line. In the model diagrams, the radial modes n = 4 and n = 5 have been connected by a black dashed line. This shift between orders to the opposite side of the diagram is not a physical discontinuity within the star; it is a geometric consequence of the module operator because of the pronounced deviation from the asymptotic regime. Regarding TIC~274211446, there is agreement between the observed and model-predicted pulsation spectra concerning the presence of three modes to the right of the first radial overtone. As the model suggests, these three modes may constitute the \ele=1 triplet of order 1, or at least two of its components, marked as $f_{110}$ and $f_{11+1}$. They are the first (f$_{1}$) and the third (f$_{3}$) peaks with highest amplitudes in the periodogram (Table~\ref{tb:Frequencies}), with a frequency difference of 0.118 \cd. Using this value together with the observed radii, R = 2.45$\pm$0.14 R$_\odot$ (Table~\ref{tb:Parameters_stars}), the result is an equatorial rotation velocity between 21 and 24 \kms. The best-fitting model has a radii of R = 2.67$\pm$0.07 R$_\odot$ and an equatorial rotation velocity of around 28$\pm$1 \kms. The observed ED shows the presence of three modes near  the first two radial modes, in particular, the 9.3 c/d mode to the left of the fundamental mode. These do not appear in the model's power spectra, possibly because they are $g$- or mixed modes, which were not calculated in the models. The observed ED does not align well with the model's ED when considering frequencies above the second order. The number of extracted frequencies and subsequently selected is insufficient to achieve a more accurate fit. However, the peak obtained in the FT (Figure~\ref{fig:dnus}) is sufficiently pronounced to lend robustness to the estimate of the large separation. Regarding TIC~16053262, there is a better agreement between the positions of the observed first radial modes and those corresponding to the models. The positions of the first overtone and the non-radial modes flanking it are in strong agreement. The peaks marked in the observed ED as $f_{11-1}$ and $f_{11+1}$ are 0.941 \cd separated. It corresponds with a rotation frequency of 0.471 \cd. Its observed radii is R = 2.29$\pm$0.08 R$_\odot$, then its equatorial rotation velocity is between 53 and 56 \kms. The best fitting model has a radii of R = 2.28$\pm$0.05 R$_\odot$ and an equatorial rotation velocity of 47$\pm$13 \kms.

\begin{figure*}   
    \centering
    \includegraphics[width=15cm,height=13cm,keepaspectratio]{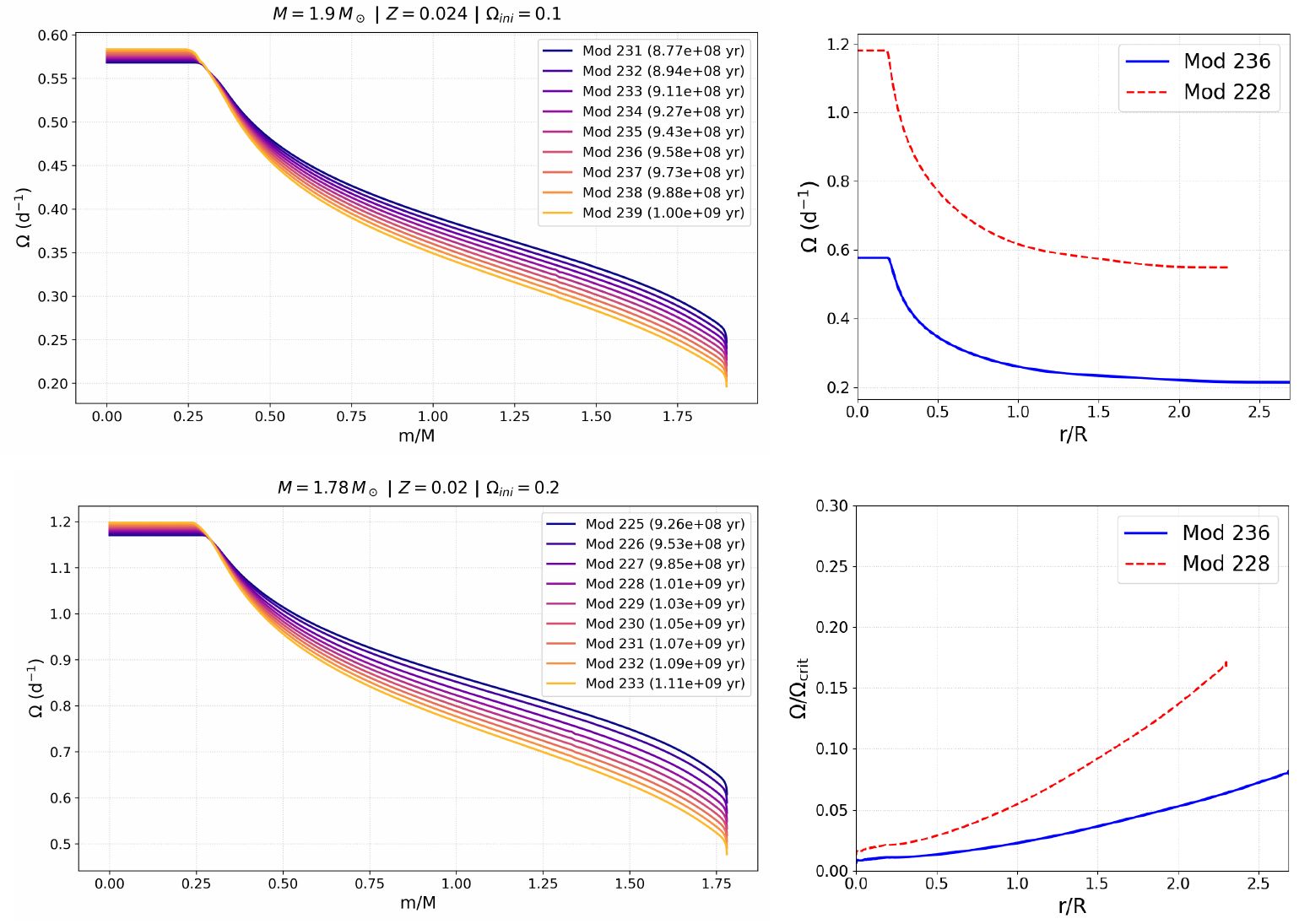}
    \caption{Left panel: Evolution of the differential rotation for the best models in the late-MS corresponding to the studied stars. Top: TIC~274211446. Bottom: TIC~16053262. Top right: Differential rotation profile of the optimal models as a function of depth. Bottom right: Profile of the ratio between the rotation frequency and the critical rotation frequency of the optimal models as a function of depth.}
    \label{fig:Dif_Rot}
\end{figure*}

The left panels of the Figure~\ref{fig:Dif_Rot} shows the evolution of the differential rotation as a function of the mass fraction for the best models corresponding to TIC~274211446 (top left) and TIC~16053262 (bottom left). In both cases, the gradient in the rotation frequency between the convective core and the radiative envelope increases as the models evolve on the MS. The right panels shows the differential rotation profile of the respective best fitting models for both stars. The gradient in the rotation frequency is larger in the model correspondig to TIC~16053262, but it never exceeds 17$\%$ of the critical velocity. This finding supports the use of the TAR approximation in \gyre\ to compute the oscillation models, allowing me to account solely for the Coriolis force and neglect the centrifugal effects that deform the star, which would otherwise require a non-perturbative treatment of rotation. Although the models are physically consistent, this differential rotation profiles should be treated with caution, as I lack observations to support them. There are no $g$-modes at low frequencies, nor enough acoustic multiplets in low radial orders, that would allow for probing the deepest interior of either star, as in the studies of \cite{Kurtz2014, Saio2015}. These researchs point to a rigid-body rotation model, with a null rotational gradient between the core and the envelope. The authors suggest a higher efficiency in the transport of angular momentum in the cases of KIC~11145123 and KIC~9244992. In these two cases, the researchers were able to use $g$- and $p$-multiplets to estimate the rotation of the core and the envelope, respectively. In this work, I did not have access to the core through the visibility of $g$-modes; consequently, the models have limitations about the physics of angular momentum transport.

\begin{figure}   
    \centering
    \includegraphics[width=9cm,height=9cm,keepaspectratio]{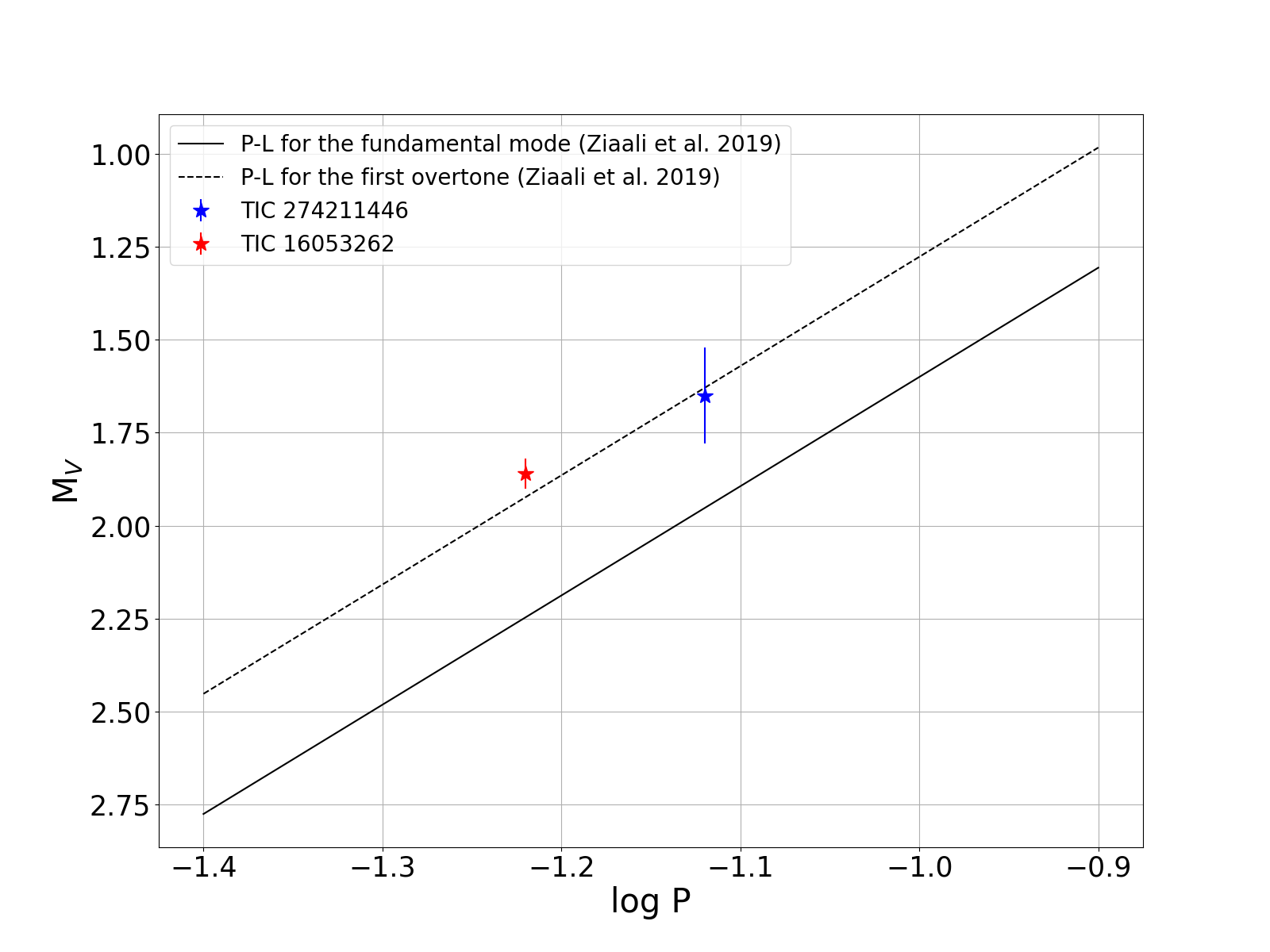}
    \caption{P-L diagram in \dss\ \citep{Ziaali2019}. Black continuous line represents the fundamental one, and the dashed line the first overtone.}
    \label{fig:PL}
\end{figure}

The Figure~\ref{fig:PL} displays the period–luminosity relation from \cite{Ziaali2019}. The corresponding magnitudes, M$_{V}$, have been calculated with the Eq~\ref{eq:Magnitude}:
\begin{equation}
    \text{M}_{\text{V}} = \text{V}\ +\ 5\log\big(\frac{\text{p}}{1000}\big)\ +\ 5\ -\ \text{A}_{V},
    \label{eq:Magnitude}
\end{equation}
where V is the apparent magnitude, A$_{\text{V}}$ = 3.1 E(B-V) is the reddening and p is the parallax in mas (Table~\ref{tb:Parameters_stars}).

In both cases, the extracted highest-amplitude frequency, f$_{1}$, lies in close proximity to the first overtone. The pulsation constant, Q, is defined by Eq.~\ref{eq:Pul_Const}:
\begin{equation}
    Q=P_{\textit{osc}} \sqrt{\frac{\rho}{\rho_{\odot}}}=\frac{1}{\text{f}_{1}} \sqrt{\frac{\rho}{\rho_{\odot}}},
    \label{eq:Pul_Const}
\end{equation}
where $P_{\textit{osc}}$=1/f$_{1}$ is the pulsation period, given in days, and $\rho$ is the mean density. Regarding TIC~274211446, f$_{1}$ = 13.05 \cd\ (Table~\ref{tb:Frequencies}) and $\rho/\rho_{\odot}$= 0.11$\pm$0.04 (Table~\ref{tb:Parameters_stars}), then $Q$=0.025 d, typical of the first radial overtone \citep{Stellingwerf1979, Fitch1981}. Regarding TIC~16053262, f$_{1}$ = 16.477 \cd\ and $\rho/\rho_{\odot}$= 0.14$\pm$0.04, resulting in $Q$=0.023 d, a value that deviates from the first radial overtone. Given that it is a Low Amplitude Delta Scuti star (LADS), which may exhibit amplitude modulation over time \citep{Bowman2016}, taking f$_{8}$ as the first radial overtone instead of f$_{1}$, the value is $Q$=0.025 d. 

\begin{figure*}   
    \centering
    \includegraphics[width=14cm,height=10cm,keepaspectratio]{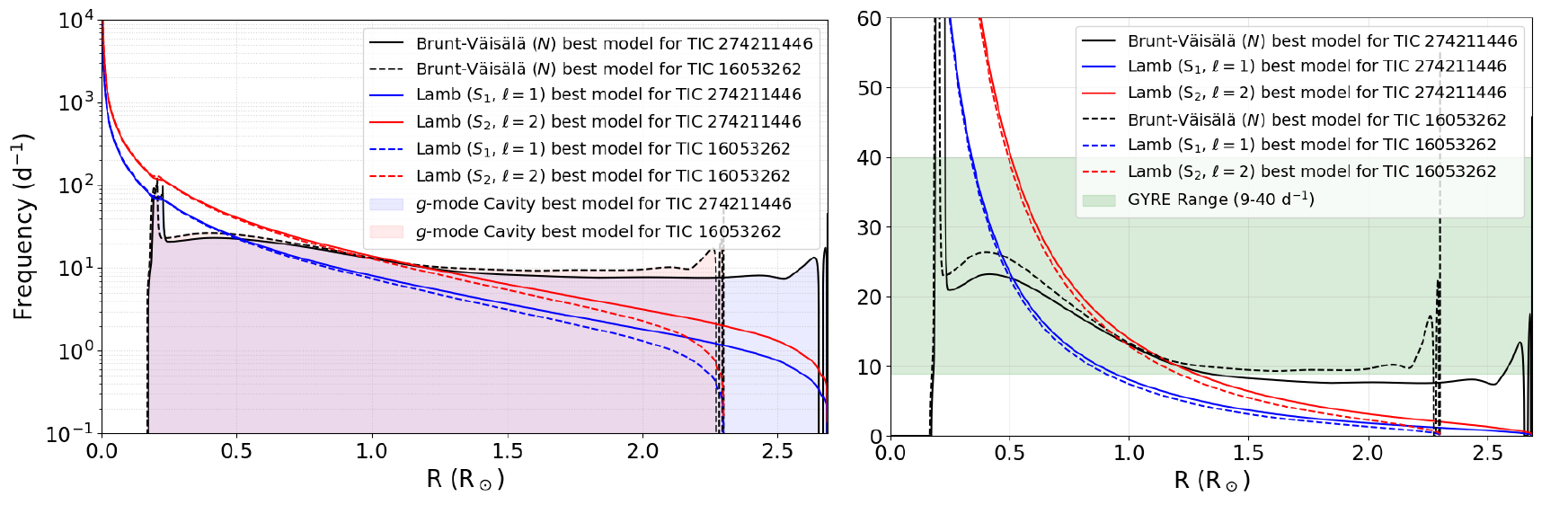}
    \caption{Left: Propagation diagrams for the non-radial modes. Right: Zoom in the scanning window of \gyre, between 9 y 40 \cd.}
    \label{fig:Propagation}
\end{figure*}

The Figure~\ref{fig:Propagation} shows what happens to the frequencies of the $p$-modes of the models as they propagate toward the interior of the star. The left panel shows the Brunt-Väisälä and the Lamb frequencies for the spherical degrees \ele=1 and \ele=2, with solid- representing TIC~274211446 and dashed-lines representing TIC~16053262. The profile of the $g$-mode cavity is very similar in both stars. The  overshooting bump at the boundary of the convective core is slightly wider for the model corresponding to TIC~274211446. The most important feature in this diagram is how the \ele=1 Lamb frequency (blue lines) penetrates the $g$-mode cavity within the frequency range scanned by \gyre. This can be seen more clearly in the expanded view of the top-right panel of the figure. The frequencies ranging 9 to 25 \cd\ are sensitive to the avoided-crossing phenomenon.

\begin{figure*}   
    \centering
    \includegraphics[width=15cm,height=13cm,keepaspectratio]{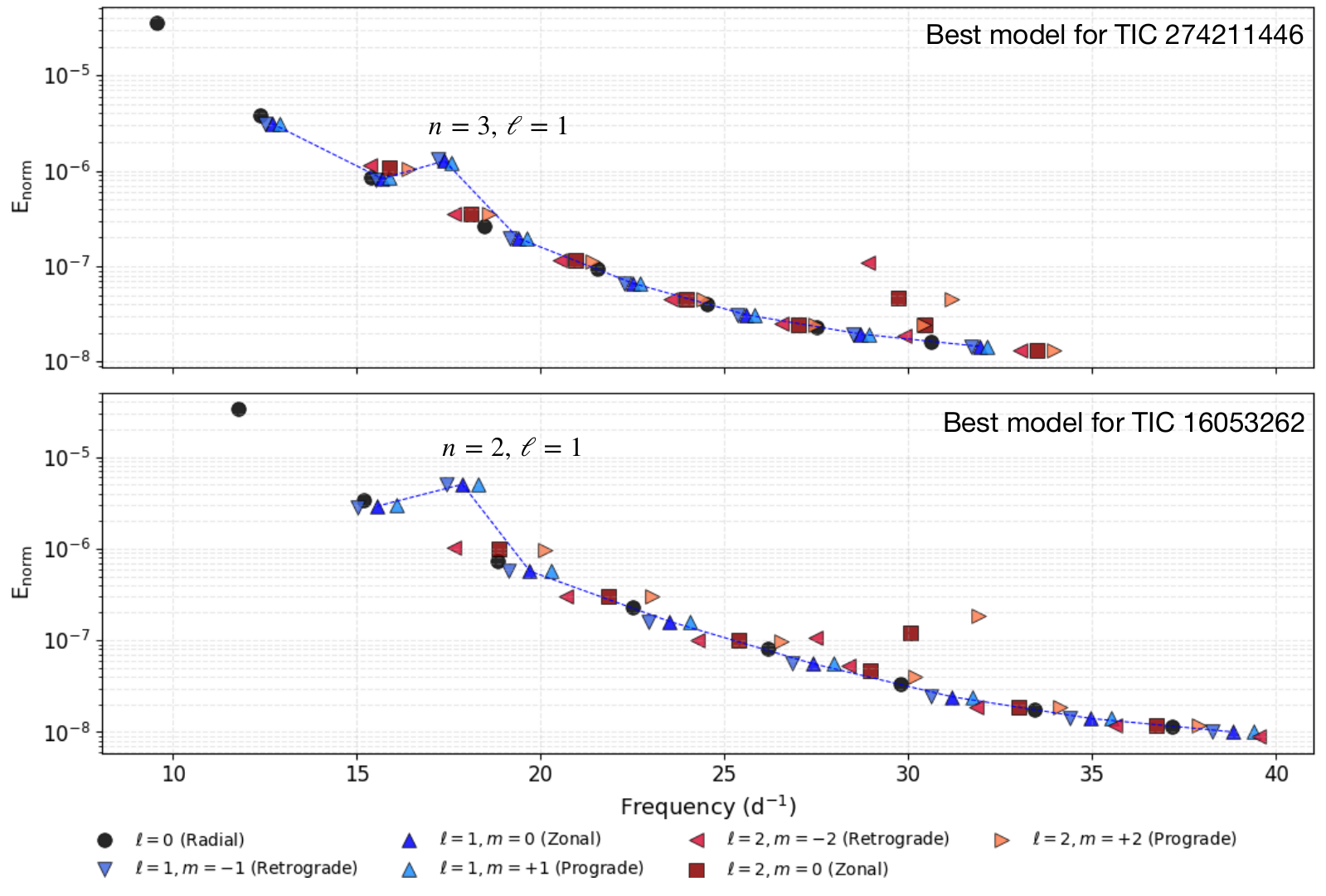}
    \caption{Inertia diagrams as a function of the frequency. The dashed blue line connects the \ele=1, $m=0$ modes.}
    \label{fig:Inertia}
\end{figure*}

The normalised inertia of a mode, E$_{norm}$, is a measure of its kinetic energy integrated over the entire star, which depends on the density of the medium, $\rho(r)$, the displacement of the mass, $\xi(r)$ and the spherical degree, \ele. The vast majority of the calculated $p$-modes follow a continuous downward trend in inertia as frequency increases (Figure~\ref{fig:Inertia}). This means that high-order modes, which are excited closer to the surface, and thus in regions of lower density, require less energy to excite. The blue dashed line connects the \ele=1, $m=0$ modes. The discontinuity observed in certain low-order \ele=1 modes reveals the hybrid nature of these frequencies. They approach the core so closely that they acquire a higher inertia value, as if it were a $g$-mode rather than a $p$-mode. No further discontinuities of this type appear in the diagram until reaching orders n= 5, 6 at spherical degree \ele=2, at frequencies around 30 \cd. Returning to the propagation diagram in the Figure~\ref{fig:Propagation}, it corresponds to values found within the overshooting bump. Unfortunately, there are not enough visible modes in the observed spectra, around the value of 30 c/d, to estimate the core frequency rotation by comparing the results with what these models predict. Furthermore, these modes have the added difficulty of being highly sensitive to the value of the overshooting. Characterizing both stars using these modes would require a grid of models with the overshooting as a free parameter. 

Except for these two discontinuities, the models are quite consistent with those of a pure \ds\ star. However, it would be necessary to calculate at least the first low order $g$-modes, as well as some mixed modes \citep{Gautam2025}, if the aim is to accurately identify the non-radial acoustic modes. This lies beyond the aims of this work, which is to reliably characterize both stars from a comparative perspective.

\begin{figure*}   
    \centering
    \includegraphics[width=16cm,height=12cm,keepaspectratio]{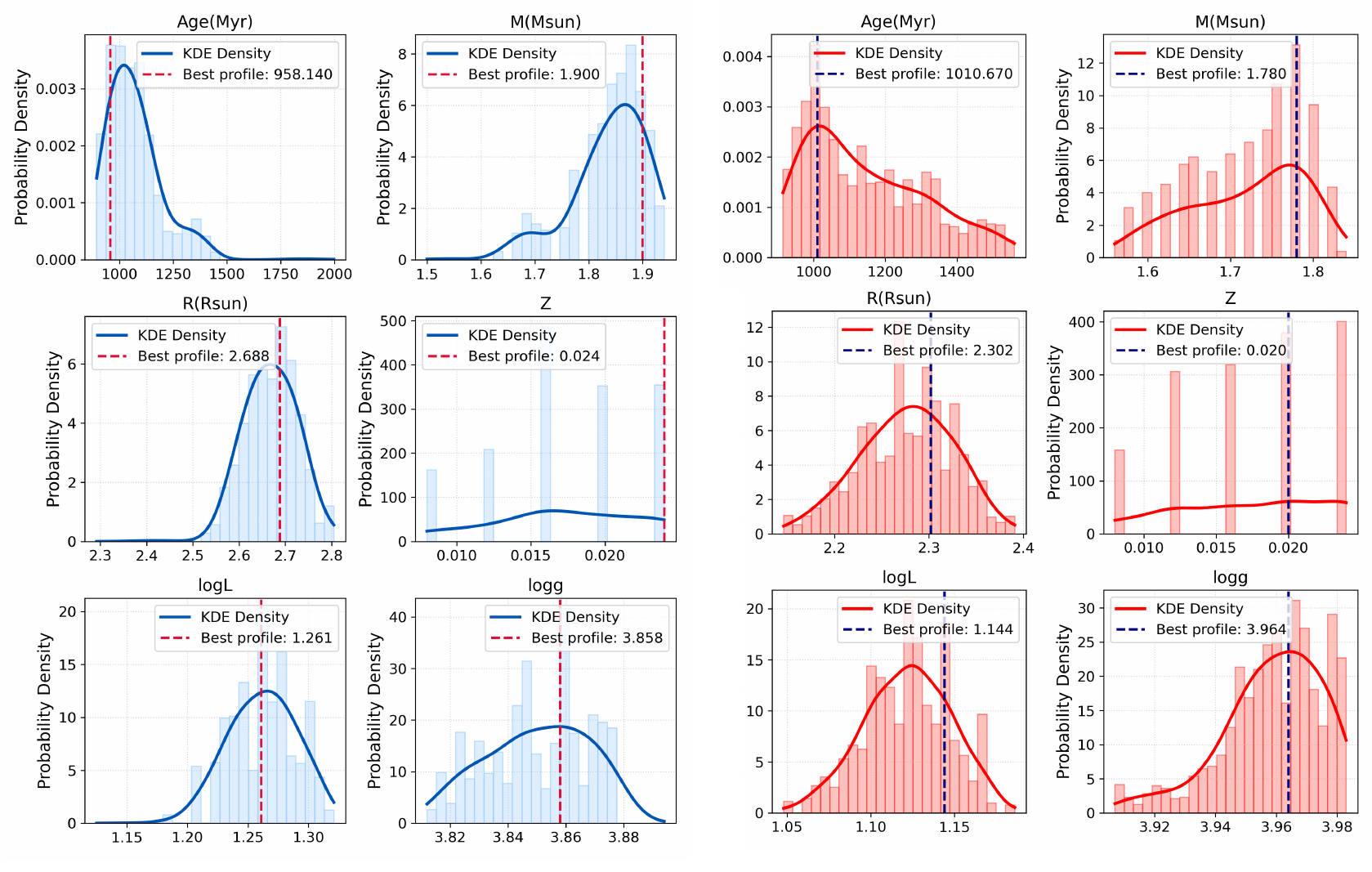}
    \caption{Histograms and KDE curves for age, mass, radius, metallicity, luminosity, and surface gravity. Left: TIC~274211446. Right: TIC~16053262.}
    \label{fig:Histograms}
\end{figure*}

Performing the statistical treatment of the models as described in Sect.~\ref{sec:grid}, I obtain the Table~\ref{tb:Parameters_models} with the weighted mean results for each parameter, accompanied by their 1-sigma uncertainties. The Figure~\ref{fig:Histograms} displays the histograms and KDE curves for the age, mass, radius, metallicity, surface gravity and luminosity. The position of the optimal model is also represented by a vertical dashed line for each parameter.

\begin{table*}
	\centering
    \caption{Parameters of the seismically constrained models with their associated $1\sigma$ uncertainties.}
    \renewcommand{\arraystretch}{1.5}
	\addtolength{\tabcolsep}{2.5 pt}
	\resizebox{\textwidth}{!}{
    \begin{tabular}{ccccccccccc}
			\hline
			TIC ID & Z & $\mathrm{log\ (L/L_{\odot})}$ & $\mathrm{M\ (M_{\odot})}$ & $\mathrm{R\ (R_{\odot})}$ & $\bar \rho\ (\bar \rho_{\odot})$ & \logg & $\mathrm{T_{eff}}$ (K) & v (\kms) & X$_{c}$ & Age (\myr) \\
			\hline
	       	274211446 & $0.017\pm0.005$ & $1.26\pm0.03$ & $1.84\pm0.07$ & $2.67\pm0.07$ & $0.14\pm0.01$ & $3.85\pm0.02$ & $7308\pm71$ & $28\pm1$ & $0.21\pm0.04$ & $1070\pm127$ \\
			16053262 & $0.017\pm0.005$ & $1.12\pm0.03$ & $1.72\pm0.07$ & $2.28\pm0.05$ & $0.21\pm0.01$ & $3.96\pm0.02$ & $7298\pm59$ & $47\pm13$ & $0.30\pm0.05$ & $1145\pm165$ \\
         
    \end{tabular}
    }

	\label{tb:Parameters_models}
\end{table*}

\section{Conclusions}\label{sec:conclusions}

This pair of stars represent two rare cases where, despite being so evolved within the MS, they display a frequency spectra clean of $g$-modes, with a minimum frequency that lies at the limit where the star can be analysed as a pure \ds\ star. A rotation ranging from slow to moderate makes it possible to characterise them using 1D-models. However, it is necessary to expand the list of calculated modes using \gyre\ to include lower order $g$- and mixed-modes, to more clearly identify additional non-radial modes. By this way, it will be possible to narrow the range of uncertainty for the model parameters.  The overall frequency analysis reveals that TIC~274211446 is more evolved than TIC~16053262, as its spectrum is slightly shifted towards lower frequencies. The former has a lower value of its core hydrogen fraction, X$_{c}$ = 0.21, indicating that its core has already contracted significantly, placing it much closer to the TAMS than TIC~116053262, with X$_{c}$ = 0.30. This one is a more compact and older star. Both stars slightly exceed 1 Gyr, reaching the limit where 1D-models can reliably characterise this type of star, according to \cite{PamosOrtega2025}. 

Determining the projected rotation speed, vsin$\textit{i}$, and therefore the inclination $\textit{i}$, would help to unravel the visibility geometry of the detected modes.  It would explain why there is greater visibility of modes in TIC~274211446 than in TIC~116053262. I think the first star is being observed from an angle between polar and equatorial, which favors the visibility of radial and non-radial modes, both zonal and sectoral, while the second may be seen from an equatorial view, where zonal modes are cancelled due to the north-south symmetry, when integrating the light over the entire disk.   

More detailed spectroscopic observations would help determine whether the atmospheres of both stars are homologous or not. In Sect.~\ref{sec:targets} I pointed out that both stars may be surrounded by a remanent circumstellar disk, which explains the intense emission in the H$_{\alpha}$ line, having ruled out a magnetic activity. The disc could produce a photometric veiling effect by equally diluting the amplitudes of all pulse modes, even obscuring those reaching the surface with less amplitude. Geometric cancellation added to this photometric veiling effect could be the reason why the frequency spectrum of TIC~116053262 is four times less dense than that of TIC~274211446. Of course, without forgetting reasons related to the efficiency of the $\kappa$-mechanism in exciting acoustic oscillations.

\section*{Data availability}
The \mesa-\gyre\ models and the complete tables containing the parameters of the modes extracted for both stars are available upon request.

\section*{Acknowledgements}
This research has made use of the SIMBAD database, operated at CDS, Strasbourg, France. This paper includes data collected by the TESS mission, which were obtained from the Mikulski Archive for Space Telescopes (MAST) at the Space Telescope Science Institute. STScI is operated by the Association of Universities for Research in Astronomy, Inc., under NASA contract NAS5-03127. Support for MAST for TESS data is provided by the NASA Office of Space Science via grant NNX16AR83G and by other grants from the New York University.

\bibliographystyle{aa}  
\bibliography{main}  

@BOOK{Aerts2010,
       author = {{Aerts}, Conny and {Christensen-Dalsgaard}, J{\o}rgen and {Kurtz}, Donald W.},
        title = "{Asteroseismology}",
         year = 2010,
        publisher = {Springer Science+Business Media B.V.},
          doi = {10.1007/978-1-4020-5803-5},
       adsurl = {https://ui.adsabs.harvard.edu/abs/2010aste.book.....A}
}

@INPROCEEDINGS{Baglin2006,
       author = {{Baglin}, A. and {Auvergne}, M. and {Barge}, P. and {Deleuil}, M. and {Catala}, C. and {Michel}, E. and {Weiss}, W. and {COROT Team}},
        title = "{Scientific Objectives for a Minisat: CoRoT}",
    booktitle = {The CoRoT Mission Pre-Launch Status - Stellar Seismology and Planet Finding},
         year = 2006,
       editor = {{Fridlund}, M. and {Baglin}, A. and {Lochard}, J. and {Conroy}, L.},
       series = {ESA Special Publication},
       volume = {1306},
        month = nov,
        pages = {33},
       adsurl = {https://ui.adsabs.harvard.edu/abs/2006ESASP1306...33B}
}

@ARTICLE{Balona2015,
       author = {{Balona}, L.~A. and {Daszy{\'n}ska-Daszkiewicz}, J. and {Pamyatnykh}, A.~A.},
        title = "{Pulsation frequency distribution in {\ensuremath{\delta}} Scuti stars}",
      journal = {Monthly Notices of Royal Astronomy Society},
         year = 2015,
        month = sep,
       volume = {452},
       number = {3},
        pages = {3073-3084},
          doi = {10.1093/mnras/stv1513},
archivePrefix = {arXiv},
       eprint = {1505.07216},
 primaryClass = {astro-ph.SR},
       adsurl = {https://ui.adsabs.harvard.edu/abs/2015MNRAS.452.3073B}
}

@ARTICLE{BF2024,
       author = {{Barcel{\'o} Forteza}, S. and {Pascual-Granado}, J. and {Su{\'a}rez}, J.~C. and {Garc{\'\i}a Hern{\'a}ndez}, A. and {Mirouh}, G.~M. and {Lares-Martiz}, M.},
        title = "{Characterizing the features of the low-amplitude peaks in delta Scuti stars with TESS}",
      journal = {Monthly Notices of Royal Astronomy Society},
         year = 2024,
        month = dec,
       volume = {535},
       number = {3},
        pages = {2189-2209},
          doi = {10.1093/mnras/stae2439},
       adsurl = {https://ui.adsabs.harvard.edu/abs/2024MNRAS.535.2189B}
}

@article{Bedding2020,
archivePrefix = {arXiv},
arxivId = {2005.06157},
author = {Bedding, Timothy R. and Murphy, Simon J. and Hey, Daniel R. and Huber, Daniel and Li, Tanda and Smalley, Barry and Stello, Dennis and White, Timothy R. and Ball, Warrick H. and Chaplin, William J. and Colman, Isabel L. and Fuller, Jim and Gaidos, Eric and Harbeck, Daniel R. and Hermes, J. J. and Holdsworth, Daniel L. and Li, Gang and Li, Yaguang and Mann, Andrew W. and Reese, Daniel R. and Sekaran, Sanjay and Yu, Jie and Antoci, Victoria and Bergmann, Christoph and Brown, Timothy M. and Howard, Andrew W. and Ireland, Michael J. and Isaacson, Howard and Jenkins, Jon M. and Kjeldsen, Hans and McCully, Curtis and Rabus, Markus and Rains, Adam D. and Ricker, George R. and Tinney, Christopher G. and Vanderspek, Roland K.},
doi = {10.1038/s41586-020-2226-8},
eprint = {2005.06157},
issn = {14764687},
journal = {Nature},
number = {7807},
pages = {147--151},
pmid = {32405022},
title = {{Very regular high-frequency pulsation modes in young intermediate-mass stars}},
volume = {581},
year = {2020}
}

@ARTICLE{Bowman2016,
       author = {{Bowman}, Dominic M. and {Kurtz}, Donald W. and {Breger}, Michel and {Murphy}, Simon J. and {Holdsworth}, Daniel L.},
        title = "{Amplitude modulation in {\ensuremath{\delta}} Sct stars: statistics from an ensemble study of Kepler targets}",
      journal = {Monthly Notices of Royal Astronomy Society},
         year = 2016,
        month = aug,
       volume = {460},
       number = {2},
        pages = {1970-1989},
          doi = {10.1093/mnras/stw1153},
archivePrefix = {arXiv},
       eprint = {1605.03955},
 primaryClass = {astro-ph.SR},
       adsurl = {https://ui.adsabs.harvard.edu/abs/2016MNRAS.460.1970B}
}

@INPROCEEDINGS{Breger2000,
       author = {{Breger}, M.},
        title = "{{\ensuremath{\delta}} Scuti stars (Review)}",
    booktitle = {Delta Scuti and Related Stars},
         year = 2000,
       editor = {{Breger}, Michel and {Montgomery}, Michael},
       series = {Astronomical Society of the Pacific Conference Series},
       volume = {210},
        month = jan,
        pages = {3},
       adsurl = {https://ui.adsabs.harvard.edu/abs/2000ASPC..210....3B}
}

@ARTICLE{Borucki2010,
       author = {{Borucki}, William J. and {Koch}, David and {Basri}, Gibor and {Batalha}, Natalie and {Brown}, Timothy and {Caldwell}, Douglas and {Caldwell}, John and {Christensen-Dalsgaard}, J{\o}rgen and {Cochran}, William D. and {DeVore}, Edna and et al.},
        title = "{Kepler Planet-Detection Mission: Introduction and First Results}",
      journal = {Science},
         year = 2010,
        month = feb,
       volume = {327},
       number = {5968},
        pages = {977},
          doi = {10.1126/science.1185402},
       adsurl = {https://ui.adsabs.harvard.edu/abs/2010Sci...327..977B}
}

@ARTICLE{Claret2019,
       author = {{Claret}, Antonio and {Torres}, Guillermo},
        title = "{The Dependence of Convective Core Overshooting on Stellar Mass: Reality Check and Additional Evidence}",
      journal = {Astrophysical Journal},
         year = 2019,
        month = may,
       volume = {876},
       number = {2},
          eid = {134},
        pages = {134},
          doi = {10.3847/1538-4357/ab1589},
archivePrefix = {arXiv},
       eprint = {1904.02714},
 primaryClass = {astro-ph.SR},
       adsurl = {https://ui.adsabs.harvard.edu/abs/2019ApJ...876..134C}
}

@ARTICLE{Dupret2005,
       author = {{Dupret}, M. -A. and {Grigahc{\`e}ne}, A. and {Garrido}, R. and {Gabriel}, M. and {Scuflaire}, R.},
        title = "{Convection-pulsation coupling. II. Excitation and stabilization mechanisms in {\ensuremath{\delta}} Sct and {\ensuremath{\gamma}} Dor stars}",
      journal = {Astronomy \& Astrophysics},
         year = 2005,
        month = jun,
       volume = {435},
       number = {3},
        pages = {927-939},
          doi = {10.1051/0004-6361:20041817},
       adsurl = {https://ui.adsabs.harvard.edu/abs/2005A&A...435..927D}
}

@INPROCEEDINGS{Dziembowski1997,
       author = {{Dziembowski}, W.},
        title = "{Sounding stellar interiors}",
    booktitle = {Sounding Solar and Stellar Interiors},
         year = 1997,
       editor = {{Provost}, Janine and {Schmider}, Francois-Xavier},
       series = {IAU Symposium},
       volume = {181},
        month = jan,
        pages = {ISBN0792348389},
       adsurl = {https://ui.adsabs.harvard.edu/abs/1997IAUS..181..317D}
}

@ARTICLE{Fitch1981,
       author = {{Fitch}, W.~S.},
        title = "{L=0, 1, 2, and 3 pulsation constants for evolutionary models of del SCT stars.}",
      journal = {Astrophysical Journal},
         year = 1981,
        month = oct,
       volume = {249},
        pages = {218-227},
          doi = {10.1086/159278},
       adsurl = {https://ui.adsabs.harvard.edu/abs/1981ApJ...249..218F}
}

@ARTICLE{Gaia2023,
       author = {{Gaia Collaboration} and {Vallenari}, A. and {Brown}, A.~G.~A. and {Prusti}, T. and {de Bruijne}, J.~H.~J. and {Arenou}, F. and {Babusiaux}, C. and {Biermann}, M. and {Creevey}, O.~L. and {Ducourant}, C. and et al.},
        title = "{Gaia Data Release 3. Summary of the content and survey properties}",
      journal = {Astronomy \& Astrophysics},
         year = 2023,
        month = jun,
       volume = {674},
          eid = {A1},
        pages = {A1},
          doi = {10.1051/0004-6361/202243940},
archivePrefix = {arXiv},
       eprint = {2208.00211},
 primaryClass = {astro-ph.GA},
       adsurl = {https://ui.adsabs.harvard.edu/abs/2023A&A...674A...1G}
}

@ARTICLE{Gautam2025,
       author = {{Gautam}, Anuj and {Murphy}, Simon J. and {Bedding}, Timothy R.},
        title = "{Modelling delta Scuti pulsations: A new grid of p, g, and f modes across pre-main-sequence to post-main-sequence evolution}",
      journal = {arXiv e-prints},
         year = 2025,
        month = jul,
          eid = {arXiv:2507.03561},
        pages = {arXiv:2507.03561},
          doi = {10.48550/arXiv.2507.03561},
archivePrefix = {arXiv},
       eprint = {2507.03561},
 primaryClass = {astro-ph.SR},
       adsurl = {https://ui.adsabs.harvard.edu/abs/2025arXiv250703561G}
}

@ARTICLE{GH2009,
       author = {{Garc{\'\i}a Hern{\'a}ndez}, A. and {Moya}, A. and {Michel}, E. and {Garrido}, R. and {Su{\'a}rez}, J.~C. and {Rodr{\'\i}guez}, E. and {Amado}, P.~J. and {Mart{\'\i}n-Ruiz}, S. and {Rolland}, A. and {Poretti}, E. and {Samadi}, R. and {Baglin}, A. and {Auvergne}, M. and {Catala}, C. and {Lefevre}, L. and {Baudin}, F.},
        title = "{Asteroseismic analysis of the CoRoT {\ensuremath{\delta}} Scuti star <ASTROBJ>HD 174936</ASTROBJ>}",
      journal = {Astronomy \& Astrophysics},
         year = 2009,
        month = oct,
       volume = {506},
       number = {1},
        pages = {79-83},
          doi = {10.1051/0004-6361/200911932},
archivePrefix = {arXiv},
       eprint = {0908.2601},
 primaryClass = {astro-ph.SR},
       adsurl = {https://ui.adsabs.harvard.edu/abs/2009A&A...506...79G}
}

@ARTICLE{GH2013,
       author = {{Garc{\'\i}a Hern{\'a}ndez}, A. and {Moya}, A. and {Michel}, E. and {Su{\'a}rez}, J.~C. and {Poretti}, E. and {Mart{\'\i}n-Ru{\'\i}z}, S. and {Amado}, P.~J. and {Garrido}, R. and {Rodr{\'\i}guez}, E. and {Rainer}, M. and {Uytterhoeven}, K. and {Rodrigo}, C. and {Solano}, E. and {Rod{\'o}n}, J.~R. and {Mathias}, P. and {Rolland}, A. and {Auvergne}, M. and {Baglin}, A. and {Baudin}, F. and {Catala}, C. and {Samadi}, R.},
        title = "{An in-depth study of HD 174966 with CoRoT photometry and HARPS spectroscopy. Large separation as a new observable for {\ensuremath{\delta}} Scuti stars}",
      journal = {Astronomy \& Astrophysics},
         year = 2013,
        month = nov,
       volume = {559},
          eid = {A63},
        pages = {A63},
          doi = {10.1051/0004-6361/201220256},
archivePrefix = {arXiv},
       eprint = {1307.2739},
 primaryClass = {astro-ph.SR},
       adsurl = {https://ui.adsabs.harvard.edu/abs/2013A&A...559A..63G}
}

@ARTICLE{GH2015,
       author = {{Garc{\'\i}a Hern{\'a}ndez}, A. and {Mart{\'\i}n-Ruiz}, S. and {Monteiro}, M{\'a}rio J.~P.~F.~G. and {Su{\'a}rez}, J.~C. and {Reese}, D.~R. and {Pascual-Granado}, J. and {Garrido}, R.},
        title = "{Observational {\ensuremath{\Delta}}{\ensuremath{\nu}}-{\ensuremath{\rho}}{\textasciimacron} Relation for {\ensuremath{\delta}} Sct Stars using Eclipsing Binaries and Space Photometry}",
      journal = {Astrophysical Journall},
         year = 2015,
        month = oct,
       volume = {811},
       number = {2},
          eid = {L29},
        pages = {L29},
          doi = {10.1088/2041-8205/811/2/L29},
archivePrefix = {arXiv},
       eprint = {1509.01111},
 primaryClass = {astro-ph.SR},
       adsurl = {https://ui.adsabs.harvard.edu/abs/2015ApJ...811L..29G}
}

@ARTICLE{GH2017,
       author = {{Garc{\'\i}a Hern{\'a}ndez}, A. and {Su{\'a}rez}, J.~C. and {Moya}, A. and {Monteiro}, M.~J.~P.~F.~G. and {Guo}, Z. and {Reese}, D.~R. and {Pascual-Granado}, J. and {Barcel{\'o} Forteza}, S. and {Mart{\'\i}n-Ruiz}, S. and {Garrido}, R. and {Nieto}, J.},
        title = "{Precise surface gravities of {\ensuremath{\delta}} Scuti stars from asteroseismology}",
      journal = {Monthly Notices of Royal Astronomy Society},
         year = 2017,
        month = oct,
       volume = {471},
       number = {1},
        pages = {L140-L144},
          doi = {10.1093/mnrasl/slx117},
archivePrefix = {arXiv},
       eprint = {1707.06835},
 primaryClass = {astro-ph.SR},
       adsurl = {https://ui.adsabs.harvard.edu/abs/2017MNRAS.471L.140G}
}

@ARTICLE{Griga2010,
       author = {{Grigahc{\`e}ne}, A. and {Antoci}, V. and {Balona}, L. and {Catanzaro}, G. and {Daszy{\'n}ska-Daszkiewicz}, J. and {Guzik}, J.~A. and {Handler}, G. and {Houdek}, G. and {Kurtz}, D.~W. and {Marconi}, M. and {Monteiro}, M.~J.~P.~F.~G. and {Moya}, A. and {Ripepi}, V. and {Su{\'a}rez}, J. -C. and {Uytterhoeven}, K. and {Borucki}, W.~J. and {Brown}, T.~M. and {Christensen-Dalsgaard}, J. and {Gilliland}, R.~L. and {Jenkins}, J.~M. and {Kjeldsen}, H. and {Koch}, D. and {Bernabei}, S. and {Bradley}, P. and {Breger}, M. and {Di Criscienzo}, M. and {Dupret}, M. -A. and {Garc{\'\i}a}, R.~A. and {Garc{\'\i}a Hern{\'a}ndez}, A. and {Jackiewicz}, J. and {Kaiser}, A. and {Lehmann}, H. and {Mart{\'\i}n-Ruiz}, S. and {Mathias}, P. and {Molenda-{\.Z}akowicz}, J. and {Nemec}, J.~M. and {Nuspl}, J. and {Papar{\'o}}, M. and {Roth}, M. and {Szab{\'o}}, R. and {Suran}, M.~D. and {Ventura}, R.},
        title = "{Hybrid {\ensuremath{\gamma}} Doradus-{\ensuremath{\delta}} Scuti Pulsators: New Insights into the Physics of the Oscillations from Kepler Observations}",
      journal = {Astrophysical Journall},
         year = 2010,
        month = apr,
       volume = {713},
       number = {2},
        pages = {L192-L197},
          doi = {10.1088/2041-8205/713/2/L192},
archivePrefix = {arXiv},
       eprint = {1001.0747},
 primaryClass = {astro-ph.SR},
       adsurl = {https://ui.adsabs.harvard.edu/abs/2010ApJ...713L.192G}
}

@ARTICLE{Guzik2021,
       author = {{Guzik}, Joyce Ann},
        title = "{Highlights of Discoveries for {\ensuremath{\delta}} Scuti Variable Stars From the Kepler Era}",
      journal = {Frontiers in Astronomy and Space Sciences},
         year = 2021,
        month = apr,
       volume = {8},
          eid = {55},
        pages = {55},
          doi = {10.3389/fspas.2021.653558},
archivePrefix = {arXiv},
       eprint = {2105.12553},
 primaryClass = {astro-ph.SR},
       adsurl = {https://ui.adsabs.harvard.edu/abs/2021FrASS...8...55G}
}

@ARTICLE{Holdsworth2014,
       author = {{Holdsworth}, Daniel L. and {Smalley}, B. and {Gillon}, M. and {Clubb}, K.~I. and {Southworth}, J. and {Maxted}, P.~F.~L. and {Anderson}, D.~R. and {Barros}, S.~C.~C. and {Collier Cameron}, A. and {Delrez}, L. and {Faedi}, F. and {Haswell}, C.~A. and {Hellier}, C. and {Horne}, K. and {Jehin}, E. and {Norton}, A.~J. and {Pollacco}, D. and {Skillen}, I. and {Smith}, A.~M.~S. and {West}, R.~G. and {Wheatley}, P.~J.},
        title = "{High-frequency A-type pulsators discovered using SuperWASP}",
      journal = {Monthly Notices of Royal Astronomy Society},
         year = 2014,
        month = apr,
       volume = {439},
       number = {2},
        pages = {2078-2095},
          doi = {10.1093/mnras/stu094},
archivePrefix = {arXiv},
       eprint = {1401.3199},
 primaryClass = {astro-ph.SR},
       adsurl = {https://ui.adsabs.harvard.edu/abs/2014MNRAS.439.2078H}
}

@ARTICLE{Kurtz2014,
       author = {{Kurtz}, Donald W. and {Saio}, Hideyuki and {Takata}, Masao and {Shibahashi}, Hiromoto and {Murphy}, Simon J. and {Sekii}, Takashi},
        title = "{Asteroseismic measurement of surface-to-core rotation in a main-sequence A star, KIC 11145123}",
      journal = {Monthly Notices of Royal Astronomy Society},
         year = 2014,
        month = oct,
       volume = {444},
       number = {1},
        pages = {102-116},
          doi = {10.1093/mnras/stu1329},
archivePrefix = {arXiv},
       eprint = {1405.0155},
 primaryClass = {astro-ph.SR},
       adsurl = {https://ui.adsabs.harvard.edu/abs/2014MNRAS.444..102K}
}

@ARTICLE{Lignieres2006,
       author = {{Ligni{\`e}res}, F. and {Rieutord}, M. and {Reese}, D.},
        title = "{Acoustic oscillations of rapidly rotating polytropic stars. I. Effects of the centrifugal distortion}",
      journal = {Astronomy \& Astrophysics},
         year = 2006,
        month = aug,
       volume = {455},
       number = {2},
        pages = {607-620},
          doi = {10.1051/0004-6361:20065015},
archivePrefix = {arXiv},
       eprint = {astro-ph/0604312},
 primaryClass = {astro-ph},
       adsurl = {https://ui.adsabs.harvard.edu/abs/2006A&A...455..607L}
}

@ARTICLE{Lindegren2021,
       author = {{Lindegren}, L. and {Klioner}, S.~A. and {Hern{\'a}ndez}, J. and {Bombrun}, A. and {Ramos-Lerate}, M. and {Steidelm{\"u}ller}, H. and {Bastian}, U. and {Biermann}, M. and {de Torres}, A. and {Gerlach}, E. and et al.},
        title = "{Gaia Early Data Release 3. The astrometric solution}",
      journal = {Astronomy \& Astrophysics},
         year = 2021,
        month = may,
       volume = {649},
          eid = {A2},
        pages = {A2},
          doi = {10.1051/0004-6361/202039709},
archivePrefix = {arXiv},
       eprint = {2012.03380},
 primaryClass = {astro-ph.IM},
       adsurl = {https://ui.adsabs.harvard.edu/abs/2021A&A...649A...2L}
}

@INCOLLECTION{Meynet2013,
       author = {{Meynet}, Georges and {Ekstrom}, Sylvia and {Maeder}, Andr{\'e} and {Eggenberger}, Patrick and {Saio}, Hideyuki and {Chomienne}, Vincent and {Haemmerl{\'e}}, Lionel},
        title = "{Models of Rotating Massive Stars: Impacts of Various Prescriptions}",
    booktitle = {Lecture Notes in Physics, Berlin Springer Verlag},
         year = 2013,
    publisher = {Lecture Notes in Physics, Volume 865. ISBN 978-3-642-33379-8. Springer-Verlag Berlin Heidelberg},
       editor = {{Goupil}, Mariejo and {Belkacem}, K{\'e}vin and {Neiner}, Coralie and {Ligni{\`e}res}, Francois and {Green}, John J.},
       volume = {865},
        pages = {3},
          doi = {10.1007/978-3-642-33380-4_1},
       adsurl = {https://ui.adsabs.harvard.edu/abs/2013LNP...865....3M}
}

@INPROCEEDINGS{Michel2017,
       author = {{Michel}, Eric and {Dupret}, Marc-Antoine and {Reese}, Daniel and {Ouazzani}, Rhita-Maria and {Debosscher}, Jonas and {Hern{\'a}ndez}, Antonio Garc{\'\i}a and {Belkacem}, Kevin and {Samadi}, Reza and {Salmon}, S{\'e}bastien and {Suarez}, Juan Carlos and {Forteza}, Sebastia Barcel{\'o}},
        title = "{What CoRoT tells us about {\ensuremath{\delta}} Scuti stars. Existence of a regular pattern and seismic indices to characterize stars}",
    booktitle = {European Physical Journal Web of Conferences},
         year = 2017,
       series = {European Physical Journal Web of Conferences},
       volume = {160},
        month = oct,
    publisher = {EDP},
          eid = {03001},
        pages = {03001},
          doi = {10.1051/epjconf/201716003001},
archivePrefix = {arXiv},
       eprint = {1705.03721},
 primaryClass = {astro-ph.SR},
       adsurl = {https://ui.adsabs.harvard.edu/abs/2017EPJWC.16003001M}
}

@ARTICLE{Mirouh2022,
       author = {{Mirouh}, Giovanni M.},
        title = "{Forward modelling and the quest for mode identification in rapidly rotating stars}",
      journal = {Frontiers in Astronomy and Space Sciences},
         year = 2022,
        month = oct,
       volume = {9},
          eid = {952296},
        pages = {952296},
          doi = {10.3389/fspas.2022.952296},
archivePrefix = {arXiv},
       eprint = {2209.09402},
 primaryClass = {astro-ph.SR},
       adsurl = {https://ui.adsabs.harvard.edu/abs/2022FrASS...9.2296M}
}

@ARTICLE{Nowakowski2003,
       author = {{Nowakowski}, Rafa{\l} M. and {Dziembowski}, Wojciech A.},
        title = "{Nonlinearity of Nonradial Modes in Evolved Stars}",
      journal = {Astrophysics and Space Science},
         year = 2003,
        month = mar,
       volume = {284},
       number = {1},
        pages = {273-276},
          doi = {10.1023/A:1023260223343},
archivePrefix = {arXiv},
       eprint = {astro-ph/0301504},
 primaryClass = {astro-ph},
       adsurl = {https://ui.adsabs.harvard.edu/abs/2003Ap&SS.284..273N}
}

@ARTICLE{PamosOrtega2022,
       author = {{Pamos Ortega}, David and {Garc{\'\i}a Hern{\'a}ndez}, Antonio and {Su{\'a}rez}, Juan Carlos and {Pascual Granado}, Javier and {Barcel{\'o} Forteza}, Sebasti{\`a} and {Rod{\'o}n}, Jos{\'e} Ram{\'o}n},
        title = "{Determining the seismic age of the young open cluster {\ensuremath{\alpha}} Per using {\ensuremath{\delta}} Scuti stars}",
      journal = {Monthly Notices of Royal Astronomy Society},
         year = 2022,
        month = jun,
       volume = {513},
       number = {1},
        pages = {374-388},
          doi = {10.1093/mnras/stac864},
archivePrefix = {arXiv},
       eprint = {2203.14256},
 primaryClass = {astro-ph.SR},
       adsurl = {https://ui.adsabs.harvard.edu/abs/2022MNRAS.513..374P}
}

@ARTICLE{PamosOrtega2023,
       author = {{Pamos Ortega}, D. and {Mirouh}, G.~M. and {Garc{\'\i}a Hern{\'a}ndez}, A. and {Su{\'a}rez Yanes}, J.~C. and {Barcel{\'o} Forteza}, S.},
        title = "{Dating young open clusters using {\ensuremath{\delta}} Scuti stars. Results for Trumpler 10 and Praesepe}",
      journal = {Astronomy \& Astrophysics},
         year = 2023,
        month = jul,
       volume = {675},
          eid = {A167},
        pages = {A167},
          doi = {10.1051/0004-6361/202346323},
archivePrefix = {arXiv},
       eprint = {2306.02791},
 primaryClass = {astro-ph.SR},
       adsurl = {https://ui.adsabs.harvard.edu/abs/2023A&A...675A.167P}
}

@ARTICLE{PamosOrtega2025,
       author = {{Pamos Ortega}, David},
        title = "{Asteroseismology of 35 Kepler and TESS {\ensuremath{\delta}} Scuti stars near the red edge of the instability strip: The limitations of {\ensuremath{\delta}} Scuti stars for dating open clusters}",
      journal = {Astronomy \& Astrophysics},
         year = 2025,
        month = oct,
       volume = {703},
          eid = {A31},
        pages = {A31},
          doi = {10.1051/0004-6361/202556586},
archivePrefix = {arXiv},
       eprint = {2509.05684},
 primaryClass = {astro-ph.SR},
       adsurl = {https://ui.adsabs.harvard.edu/abs/2025A&A...703A..31P}
}

@ARTICLE{Paparo2016,
       author = {{Papar{\'o}}, M. and {Benk{\H{o}}}, J.~M. and {Hareter}, M. and {Guzik}, J.~A.},
        title = "{Unexpected Series of Regular Frequency Spacing of {\ensuremath{\delta}} Scuti Stars in the Non-asymptotic Regime. I. The Methodology}",
      journal = {Astrophysical Journal},
         year = 2016,
        month = may,
       volume = {822},
       number = {2},
          eid = {100},
        pages = {100},
          doi = {10.3847/0004-637X/822/2/100},
archivePrefix = {arXiv},
       eprint = {1603.02050},
 primaryClass = {astro-ph.SR},
       adsurl = {https://ui.adsabs.harvard.edu/abs/2016ApJ...822..100P}
}

@ARTICLE{Paxton2011,
       author = {{Paxton}, Bill and {Bildsten}, Lars and {Dotter}, Aaron and {Herwig}, Falk and {Lesaffre}, Pierre and {Timmes}, Frank},
        title = "{Modules for Experiments in Stellar Astrophysics (MESA)}",
      journal = {Astrophysical Journals},
         year = 2011,
        month = jan,
       volume = {192},
       number = {1},
          eid = {3},
        pages = {3},
          doi = {10.1088/0067-0049/192/1/3},
archivePrefix = {arXiv},
       eprint = {1009.1622},
 primaryClass = {astro-ph.SR},
       adsurl = {https://ui.adsabs.harvard.edu/abs/2011ApJS..192....3P}
}

@ARTICLE{Paxton2013,
       author = {{Paxton}, Bill and {Cantiello}, Matteo and {Arras}, Phil and {Bildsten}, Lars and {Brown}, Edward F. and {Dotter}, Aaron and {Mankovich}, Christopher and {Montgomery}, M.~H. and {Stello}, Dennis and {Timmes}, F.~X. and et al.},
        title = "{Modules for Experiments in Stellar Astrophysics (MESA): Planets, Oscillations, Rotation, and Massive Stars}",
      journal = {Astrophysical Journals},
         year = 2013,
        month = sep,
       volume = {208},
       number = {1},
          eid = {4},
        pages = {4},
          doi = {10.1088/0067-0049/208/1/4},
archivePrefix = {arXiv},
       eprint = {1301.0319},
 primaryClass = {astro-ph.SR},
       adsurl = {https://ui.adsabs.harvard.edu/abs/2013ApJS..208....4P}
}

@MISC{Paxton2019,
       author = {{Paxton}, Bill},
        title = "{Modules for Experiments in Stellar Astrophysics (MESA)}",
         year = 2019,
        month = sep,
          eid = {10.5281/zenodo.3473377},
          doi = {10.5281/zenodo.3473377},
      version = {r12115},
    publisher = {Zenodo},
       adsurl = {https://ui.adsabs.harvard.edu/abs/2019zndo...3473377P}
}

@ARTICLE{RBallesta2021,
       author = {{Ram{\'o}n-Ballesta}, A. and {Garc{\'\i}a Hern{\'a}ndez}, A. and {Su{\'a}rez}, J.~C. and {Rod{\'o}n}, J.~R. and {Pascual-Granado}, J. and {Garrido}, R.},
        title = "{Study of rotational splittings in {\ensuremath{\delta}} Scuti stars using pattern finding techniques}",
      journal = {Monthly Notices of Royal Astronomy Society},
         year = 2021,
        month = aug,
       volume = {505},
       number = {4},
        pages = {6217-6224},
          doi = {10.1093/mnras/stab1719},
archivePrefix = {arXiv},
       eprint = {2106.10059},
 primaryClass = {astro-ph.SR},
       adsurl = {https://ui.adsabs.harvard.edu/abs/2021MNRAS.505.6217R}
}

@INPROCEEDINGS{Ricker,
       author = {{Ricker}, George R. and {Winn}, Joshua N. and {Vanderspek}, Roland and
         {Latham}, David W. and {Bakos}, G{\'a}sp{\'a}r. {\'A}. and
         {Bean}, Jacob L. and {Berta-Thompson}, Zachory K. and
         {Brown}, Timothy M. and {Buchhave}, Lars and {Butler}, Nathaniel R. and
         {Butler}, R. Paul and {Chaplin}, William J. and {Charbonneau}, David and
         {Christensen-Dalsgaard}, J{\o}rgen and {Clampin}, Mark and
         {Deming}, Drake and {Doty}, John and {De Lee}, Nathan and
         {Dressing}, Courtney and {Dunham}, E.~W. and {Endl}, Michael and
         {Fressin}, Francois and {Ge}, Jian and {Henning}, Thomas and
         {Holman}, Matthew J. and {Howard}, Andrew W. and {Ida}, Shigeru and
         {Jenkins}, Jon and {Jernigan}, Garrett and {Johnson}, John A. and
         {Kaltenegger}, Lisa and {Kawai}, Nobuyuki and {Kjeldsen}, Hans and
         {Laughlin}, Gregory and {Levine}, Alan M. and {Lin}, Douglas and
         {Lissauer}, Jack J. and {MacQueen}, Phillip and {Marcy}, Geoffrey and
         {McCullough}, P.~R. and {Morton}, Timothy D. and {Narita}, Norio and
         {Paegert}, Martin and {Palle}, Enric and {Pepe}, Francesco and
         {Pepper}, Joshua and {Quirrenbach}, Andreas and {Rinehart}, S.~A. and
         {Sasselov}, Dimitar and {Sato}, Bun'ei and {Seager}, Sara and
         {Sozzetti}, Alessandro and {Stassun}, Keivan G. and {Sullivan}, Peter and
         {Szentgyorgyi}, Andrew and {Torres}, Guillermo and {Udry}, Stephane and
         {Villasenor}, Joel},
        title = "{Transiting Exoplanet Survey Satellite (TESS)}",
    booktitle = {Proceedings of the SPIE},
         year = 2014,
       series = {Society of Photo-Optical Instrumentation Engineers (SPIE) Conference Series},
       volume = {9143},
        month = aug,
          eid = {914320},
        pages = {914320},
          doi = {10.1117/12.2063489},
archivePrefix = {arXiv},
       eprint = {1406.0151},
 primaryClass = {astro-ph.EP},
       adsurl = {https://ui.adsabs.harvard.edu/abs/2014SPIE.9143E..20R}
}

@ARTICLE{RodriguezBreger2001,
       author = {{Rodr{\'\i}guez}, E. and {Breger}, M.},
        title = "{delta Scuti and related stars: Analysis of the R00 Catalogue}",
      journal = {Astronomy \& Astrophysics},
         year = 2001,
        month = jan,
       volume = {366},
        pages = {178-196},
          doi = {10.1051/0004-6361:20000205},
       adsurl = {https://ui.adsabs.harvard.edu/abs/2001A&A...366..178R}
}

@ARTICLE{Royer2007,
       author = {{Royer}, F. and {Zorec}, J. and {G{\'o}mez}, A.~E.},
        title = "{Rotational velocities of A-type stars. III. Velocity distributions}",
      journal = {Astronomy \& Astrophysics},
         year = 2007,
        month = feb,
       volume = {463},
       number = {2},
        pages = {671-682},
          doi = {10.1051/0004-6361:20065224},
archivePrefix = {arXiv},
       eprint = {astro-ph/0610785},
 primaryClass = {astro-ph},
       adsurl = {https://ui.adsabs.harvard.edu/abs/2007A&A...463..671R}
}

@ARTICLE{Saio2015,
       author = {{Saio}, Hideyuki and {Kurtz}, Donald W. and {Takata}, Masao and {Shibahashi}, Hiromoto and {Murphy}, Simon J. and {Sekii}, Takashi and {Bedding}, Timothy R.},
        title = "{Asteroseismic measurement of slow, nearly uniform surface-to-core rotation in the main-sequence F star KIC 9244992}",
      journal = {Monthly Notices of Royal Astronomy Society},
         year = 2015,
        month = mar,
       volume = {447},
       number = {4},
        pages = {3264-3277},
          doi = {10.1093/mnras/stu2696},
archivePrefix = {arXiv},
       eprint = {1412.5362},
 primaryClass = {astro-ph.SR},
       adsurl = {https://ui.adsabs.harvard.edu/abs/2015MNRAS.447.3264S}
}

@ARTICLE{Stassun2019,
       author = {{Stassun}, K.~G.},
        title = "{VizieR Online Data Catalog: TESS Input Catalog - v8.0 (TIC-8) (Stassun+, 2019)}",
      journal = {VizieR Online Data Catalog},
         year = 2019,
        month = nov,
          eid = {IV/38},
        pages = {IV/38},
       adsurl = {https://ui.adsabs.harvard.edu/abs/2019yCat.4038....0S}
}

@ARTICLE{Stellingwerf1979,
       author = {{Stellingwerf}, R.~F.},
        title = "{Pulsation in the lower Cepheid strip. I. Linear survey.}",
      journal = {Astrophysical Journal},
         year = 1979,
        month = feb,
       volume = {227},
        pages = {935-942},
          doi = {10.1086/156802},
       adsurl = {https://ui.adsabs.harvard.edu/abs/1979ApJ...227..935S}
}

@ARTICLE{Suarez2014,
       author = {{Su{\'a}rez}, J.~C. and {Garc{\'\i}a Hern{\'a}ndez}, A. and {Moya}, A. and {Rodrigo}, C. and {Solano}, E. and {Garrido}, R. and {Rod{\'o}n}, J.~R.},
        title = "{Measuring mean densities of {\ensuremath{\delta}} Scuti stars with asteroseismology. Theoretical properties of large separations using TOUCAN}",
      journal = {Astronomy \& Astrophysics},
         year = 2014,
        month = mar,
       volume = {563},
          eid = {A7},
        pages = {A7},
          doi = {10.1051/0004-6361/201322270},
archivePrefix = {arXiv},
       eprint = {1402.0723},
 primaryClass = {astro-ph.SR},
       adsurl = {https://ui.adsabs.harvard.edu/abs/2014A&A...563A...7S}
}

@ARTICLE{Tassoul1980,
       author = {{Tassoul}, M.},
        title = "{Asymptotic approximations for stellar nonradial pulsations.}",
      journal = {Astrophysical Journals},
         year = 1980,
        month = aug,
       volume = {43},
        pages = {469-490},
          doi = {10.1086/190678},
       adsurl = {https://ui.adsabs.harvard.edu/abs/1980ApJS...43..469T}
}

@ARTICLE{Trilling2008,
       author = {{Trilling}, D.~E. and {Bryden}, G. and {Beichman}, C.~A. and {Rieke}, G.~H. and {Su}, K.~Y.~L. and {Stansberry}, J.~A. and {Blaylock}, M. and {Stapelfeldt}, K.~R. and {Beeman}, J.~W. and {Haller}, E.~E.},
        title = "{Debris Disks around Sun-like Stars}",
      journal = {Astrophysical Journal},
         year = 2008,
        month = feb,
       volume = {674},
       number = {2},
        pages = {1086-1105},
          doi = {10.1086/525514},
archivePrefix = {arXiv},
       eprint = {0710.5498},
 primaryClass = {astro-ph},
       adsurl = {https://ui.adsabs.harvard.edu/abs/2008ApJ...674.1086T}
}

@ARTICLE{Townsend2013,
       author = {{Townsend}, R.~H.~D. and {Teitler}, S.~A.},
        title = "{GYRE: an open-source stellar oscillation code based on a new Magnus Multiple Shooting scheme}",
      journal = {Monthly Notices of Royal Astronomy Society},
         year = 2013,
        month = nov,
       volume = {435},
       number = {4},
        pages = {3406-3418},
          doi = {10.1093/mnras/stt1533},
archivePrefix = {arXiv},
       eprint = {1308.2965},
 primaryClass = {astro-ph.SR},
       adsurl = {https://ui.adsabs.harvard.edu/abs/2013MNRAS.435.3406T}
}

@ARTICLE{Uytterhoeven2011,
       author = {{Uytterhoeven}, K. and {Moya}, A. and {Grigahc{\`e}ne}, A. and {Guzik}, J.~A. and {Guti{\'e}rrez-Soto}, J. and {Smalley}, B. and {Handler}, G. and {Balona}, L.~A. and {Niemczura}, E. and {Fox Machado}, L. and {Benatti}, S. and {Chapellier}, E. and {Tkachenko}, A. and {Szab{\'o}}, R. and {Su{\'a}rez}, J.~C. and {Ripepi}, V. and {Pascual}, J. and {Mathias}, P. and {Mart{\'\i}n-Ru{\'\i}z}, S. and {Lehmann}, H. and {Jackiewicz}, J. and {Hekker}, S. and {Gruberbauer}, M. and {Garc{\'\i}a}, R.~A. and {Dumusque}, X. and {D{\'\i}az-Fraile}, D. and {Bradley}, P. and {Antoci}, V. and {Roth}, M. and {Leroy}, B. and {Murphy}, S.~J. and {De Cat}, P. and {Cuypers}, J. and {Kjeldsen}, H. and {Christensen-Dalsgaard}, J. and {Breger}, M. and {Pigulski}, A. and {Kiss}, L.~L. and {Still}, M. and {Thompson}, S.~E. and {van Cleve}, J.},
        title = "{The Kepler characterization of the variability among A- and F-type stars. I. General overview}",
      journal = {Astronomy \& Astrophysics},
         year = 2011,
        month = oct,
       volume = {534},
          eid = {A125},
        pages = {A125},
          doi = {10.1051/0004-6361/201117368},
archivePrefix = {arXiv},
       eprint = {1107.0335},
 primaryClass = {astro-ph.SR},
       adsurl = {https://ui.adsabs.harvard.edu/abs/2011A&A...534A.125U}
}

@ARTICLE{Zhang2023,
       author = {{Zhang}, Li-yun and {Su}, Tianhao and {Misra}, Prabhakar and {Han}, Xianming L. and {Meng}, Gang and {Pi}, Qingfeng and {Yang}, Jiawei},
        title = "{Stellar Parameters and Spectroscopic Properties of TESS Objects Observed in the LAMOST Low- and Medium-resolution Spectral Survey}",
      journal = {Astrophysical Journals},
         year = 2023,
        month = jan,
       volume = {264},
       number = {1},
          eid = {17},
        pages = {17},
          doi = {10.3847/1538-4365/ac9b28},
       adsurl = {https://ui.adsabs.harvard.edu/abs/2023ApJS..264...17Z}
}

@ARTICLE{Zhao2024,
       author = {{Guo}, Zhao and {Bedding}, Timothy R. and {Pamyatnykh}, A.~A. and {Kurtz}, Donald W. and {Li}, Gang and {Gautam}, Anuj and {Murphy}, Simon J. and {Aerts}, Conny},
        title = "{Oscillation frequencies of moderately rotating delta scuti stars: asymmetric mode splittings due to non-spherical distortion}",
      journal = {Monthly Notices of Royal Astronomy Society},
         year = 2024,
        month = dec,
       volume = {535},
       number = {4},
        pages = {2927-2938},
          doi = {10.1093/mnras/stae2423},
archivePrefix = {arXiv},
       eprint = {2406.15678},
 primaryClass = {astro-ph.SR},
       adsurl = {https://ui.adsabs.harvard.edu/abs/2024MNRAS.535.2927G}
}

@ARTICLE{Ziaali2019,
       author = {{Ziaali}, Elham and {Bedding}, Timothy R. and {Murphy}, Simon J. and {Van Reeth}, Timothy and {Hey}, Daniel R.},
        title = "{The period-luminosity relation for {\ensuremath{\delta}} Scuti stars using Gaia DR2 parallaxes}",
      journal = {Monthly Notices of Royal Astronomy Society},
         year = 2019,
        month = jul,
       volume = {486},
       number = {3},
        pages = {4348-4353},
          doi = {10.1093/mnras/stz1110},
archivePrefix = {arXiv},
       eprint = {1904.08101},
 primaryClass = {astro-ph.SR},
       adsurl = {https://ui.adsabs.harvard.edu/abs/2019MNRAS.486.4348Z}
}

\clearpage

\onecolumn
\begin{appendices}

\setcounter{table}{0} 
\renewcommand{\thetable}{A.\arabic{table}}

\section{First 35 extracted frequencies of both stars}
\label{AppendixA}

\begin{table*}[h!]
\centering
    
	\caption{Frequencies, amplitudes, phases, SNRs and some identified modes of the first 35 extracted frequencies of both stars.}
    \label{tb:Frequencies}
    \renewcommand{\arraystretch}{1.5}
	\addtolength{\tabcolsep}{2.2 pt}
	\resizebox{\textwidth}{!}{
    \begin{tabular}{cccccc|cccccc}
    
			\hline
			\multicolumn{6}{c|}{\textbf{TIC~274211446 }} & \multicolumn{6}{c}{\textbf{TIC~16053262}} \\
			\hline
            
            ID & f(\cd) & A(mmag) & $\phi$ & SNR & Mode & ID & f(\cd) & A(mmag) & $\phi$ & SNR & Mode \\
            \hline
            f$_{1}$ & 13.050  & 2.59 & 0.42 & 157.80 & &f$_{1}$ & 16.477 & 1.49 & 0.65 & 177.46 & \\
            f$_{2}$ & 9.601 &	1.78 &	0.28 &	112.98 & $f_{100}$ & f$_{2}$ & 21.001 &	1.34 &	0.98 &	171.16 & \\
            f$_{3}$ & 12.868 & 1.64 &	0.14 &	101.51	& &	f$_{3}$ & 17.706 & 1.10 &	0.27 &	137.21 & \\
            f$_{4}$ & 11.839 & 1.62 & 0.83 &	101.20	& &	f$_{4}$ & 18.232 & 0.83 &	0.60 &	102.99 & \\
            f$_{5}$ & 22.235 & 1.09 &	0.66 &	67.14 & & f$_{5}$ & 11.715 &	0.72 &	0.09 &	91.08 &	$f_{100}$ \\
            f$_{6}$ & 24.241 & 1.05 &	0.64 &	65.96 &	$f_{600}$ &	f$_{6}$ & 13.977 & 0.84 & 1.00 &	90.33 & \\
            f$_{7}$ & 17.848 & 0.96 &	0.82 &	59.72 &	6 \cd\ alias of f$_{4}$ & f$_{7}$ & 22.243 & 0.58 &	0.72 &	71.74 &\\
            f$_{8}$ & 20.968 &	0.88 &	0.33 &	55.19 & &	f$_{8}$ & 15.152 &	0.53 &	0.73 &	66.24 &	$f_{200}$ \\
            f$_{9}$ & 9.312 &	0.85 &	0.61 &	51.68 && f$_{9}$ & 16.349	& 0.46 &	0.22 &	55.55 &	\\
            f$_{10}$ & 11.094 &	0.75 &	0.14 &	47.01 && f$_{10}$ & 19.729 & 	0.37 &	0.53 &	46.32 & \\
            f$_{11}$ & 11.212 &	0.70 &	0.36 &	42.73 &	-f$_{1}$ + f$_{6}$ & f$_{11}$ & 21.046	& 0.38 &	0.73 &	45.01 & \\
            f$_{12}$ & 18.673 &	0.66 &	0.76 &	42.57 &&	f$_{12}$ &	16.860 &	0.36 &	0.84 &	44.78	&\\
            f$_{13}$ & 15.927 &	0.62 &	0.99 &	38.30 &&	f$_{13}$ & 10.526 &	0.34 &	0.25 &	42.86 &	-f$_{5}$ + f$_{7}$ \\
            f$_{14}$ & 15.823 &	0.50 &	0.57 &	31.26 &&	f$_{14}$ &  29.609 &	0.26 &	0.97 &	32.15 &\\
            f$_{15}$ & 17.331 &	0.50 &	0.85 &	31.13 &&	f$_{15}$ & 25.319 &	0.21 &	0.09 &	26.22 &\\
            f$_{16}$ & 12.782 &	0.51 &	0.22 &	30.48 &&	f$_{16}$ & 22.057 &	0.20 &	0.72 &	25.14 & \\
            f$_{17}$ & 18.442 &	0.49 &	0.37 &	30.45 &	$f_{400}$ & f$_{17}$ & 16.742	& 0.18 &	0.21 &	21.96 &\\
            f$_{18}$ & 22.671 &	0.49 &	0.41 &	30.33 &	f$_{1}$ + f$_{9}$ & f$_{18}$ &	15.673 &	0.17 &	0.43 &	20.27 &	-f$_{6}$ + f$_{14}$\\
            f$_{19}$ & 26.819 &	0.47 &	0.50 &	28.86 &&	f$_{19}$ & 12.187 &	0.16 &	0.81 &	19.88 & \\
            f$_{20}$ & 21.422 &	0.44 &	0.22 &	27.85 &	f$_{2}$ + f$_{4}$ & f$_{20}$ &	16.079 &	0.11 &	0.74 &	13.88 &\\
            f$_{21}$ & 13.638 &	0.41 &	0.54 &	25.64 && 	f$_{21}$ &	26.312 &	0.11 &	0.51 &	13.83	&\\
            f$_{22}$ & 18.129 &	0.41 &	0.11 &	25.59 &&	f$_{22}$ &	15.817 &	0.11 &	0.37 &	13.31 &\\
            f$_{23}$ & 23.848 &	0.37 &	0.84 &	23.13 &	f$_{10}$ + f$_{16}$ & f$_{23}$ &	25.021 &	0.10 &	0.73 &	12.74 &\\
            f$_{24}$ & 15.460 &	0.34 &	0.03 &	22.96 &&		f$_{24}$ & 20.049 &	0.09 &	0.40 &	11.82	 &\\
            f$_{25}$ & 24.726 &	0.36 &	0.59 &	22.50 &	f$_{3}$ + f$_{4}$	& f$_{25}$ & 16.594 &	0.09 &	0.53 &	11.68 &\\
            f$_{26}$ & 10.839 &	0.35 &	0.16 &	22.23 &	1 \cd\ alias of f$_{4}$ & f$_{26}$ &	1.447 &	0.09 &	0.79 &	10.76 &\\	
            f$_{27}$ & 12.091 &	0.35 &	0.82 &	21.95 &	Modulation of f$_{4}$ & f$_{27}$ & 18.024 &	0.08 &	0.89 &	10.62 &\\
            f$_{28}$ & 27.974 &	0.34 &	0.47 &	21.52 &	f$_{9}$ + f$_{12}$ & f$_{28}$ & 18.766	& 0.08 &	0.50 &	9.96 &	$f_{300}$ \\
            f$_{29}$ & 12.574 &	0.35 &	0.21 &	21.50 &	$f_{200}$ & f$_{29}$ & 3.384	& 0.08 &	0.52 &	9.48 &\\
            f$_{30}$ & 14.246 &	0.34 &	0.37 &	21.20 &	f$_{19}$ - f$_{29}$ & f$_{30}$ & 20.551 &	0.08 &	0.62 &	9.34 &\\
            f$_{31}$ & 15.068 &	0.34 &	0.75 &	20.26 &&	f$_{31}$ &	13.865 &	0.07 &	0.63 &	9.08 &\\
            f$_{32}$ & 15.255 &	0.32 &	0.28 &	19.69 &	$f_{300}$ & f$_{32}$ &	21.222 &	0.07 &	0.97 &	8.86 &\\
            f$_{33}$ & 26.376 &	0.30 &	0.64 &	18.59 &	f$_{10}$ + f$_{32}$ & f$_{33}$ &	14.876 &	0.06 &	0.67 &	7.94 &\\
            f$_{34}$ & 19.285 &	0.31 &	0.49 &	18.38 && f$_{34}$ & 12.505 &	0.06 &	0.14 &	7.61	&\\
            f$_{35}$ & 13.304 &	0.29 &	0.10 &	18.00 && f$_{35}$ &	24.257	& 0.06 &	0.30 &	7.53	&\\      
            
    \end{tabular}
    }

\end{table*}

\end{appendices}
\end{document}